# $\mathcal{PT}-$symmetric cross-injection dual optoelectronic oscillator

Mehedi Hasan[1*], Abhijit Banerjee[2] & Trevor Hall[1]

[1]University of Ottawa, Electronic Engineering & Computer Science, Advanced Research Complex, 25 Templeton St, Ottawa, Ontario, K1N 6N5, Canada

[2]Academy of Technology, Department of Electronics & Communications Engineering, Adisaptagram, Hooghly, West Bengal, 712121, India.

*mhasa067@uottawa.ca

## Abstract

An optoelectronic oscillator (OEO) is a time delay oscillator (TDO) that uses photonics technology to provide the long delay required to generate pristine microwave carriers. Parity-time ($\mathcal{PT}$) symmetry concepts applied to an OEO offer the potential to achieve combined low phase noise and high sidemode suppression. A TDO composed of a pair of identical ring resonators coupled by a $2 \times 2$ coupler is modelled, and the coupler transmission matrix required for the oscillator to be $\mathcal{PT}-$ symmetric is derived. In a first configuration, the coupler is interpreted as the composition of a gain/loss block and a Mach-Zehnder interferometer (MZI) block. In practice, there are excess losses that must be compensated by a special dual amplifier with saturation characteristics compatible with $\mathcal{PT}-$symmetry. The $\mathcal{PT}-$ symmetry phase transition determined by the gain/loss and the MZI differential phase parameters is found to be global and not local in its effect on modes. This is resolved by placing a short delay line within one arm of the MZI resulting in a frequency dependent and hence local mode-selective $\mathcal{PT}-$ symmetry phase transition. In addition, it is demonstrated that the first configuration may be transformed into a second but equivalent configuration as a cross-injection dual TDO with imbalanced delays. The local $\mathcal{PT}-$ symmetry phase transition may then be understood in terms of the Vernier effect. The cross-injection perspective facilitates the extension of the theory to a phase-only model of an established oscillation that inter alia provides an analytic prediction of the phase-noise spectral density. Advantageously, the second configuration enables the special dual amplifier to be replaced by a pair of matched but otherwise independent amplifiers. Thereby, the second configuration is amenable to practical implementation as a dual OEO using standard RF-photonic and RF-electronic components. The theory is validated by complex envelope model simulations using Simulink™ and phase model analytic results evaluated using MATLAB™. There is excellent agreement between the theoretical and simulation results.

## 1. Introduction

An OEO is a TDO that leverages photonics technology to provide the long delay required to generate pristine microwave carriers [1]. The low attenuation of standard single mode optical fibre $\sim 0.3 \ dB/km$ permits optical fibre delay lines of many kilometres to be used enabling exceptionally low phase-noise while incurring losses that can be readily compensated with available RF amplifier technology. However, a TDO supports a multitude of modes regularly spaced in frequency by the reciprocal of the delay time which create substantial spurious sidemode resonances in the phase-noise spectrum. An optical fibre delay line $5 \ km$ in length has a delay time of $25 \ \mu s$ resulting in a $40 \ kHz$ frequency interval between modes. At $10 \ GHz$ a high-Q dielectric resonator can offer a $-3 \ dB$ passband of $3 \ MHz$ within which 75

# $\mathcal{PT}$−symmetric cross-injection dual optoelectronic oscillator

of these potential oscillating modes would fall. With a wider and/or flatter passband sustained multimode oscillation can occur. The application of $\mathcal{PT}$ symmetry to OEOs of has attracted attention as a potential solution to the achievement of combined low-phase noise and high sidemode suppression.

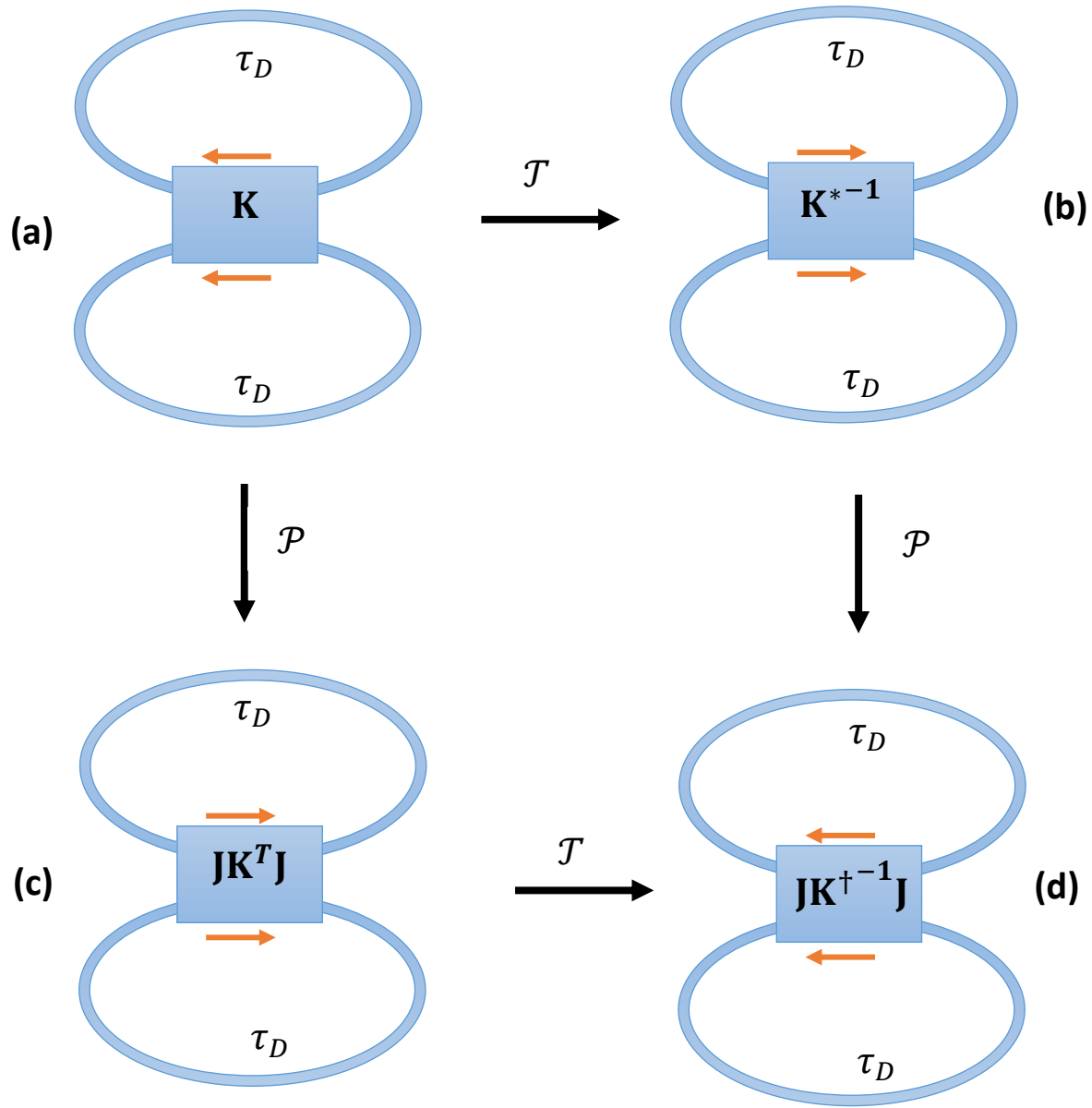

Figure 1. An illustration of the commutation of the parity $\mathcal{P}$ and time $\mathcal{T}$ symmetry transformations applied to a coupled dual time delay oscillator. (a) The oscillator is composed of a pair of identical delay lines with delay time $\tau_D$ which are closed into a coupled pair of ring oscillators by the active coupler represented by a rectangular block. The direction of propagation of the oscillations is shown by the red arrows. The coupler is characterised in the Fourier domain by a transmission matrix $\mathbf{K}$. (b) The transformation of the oscillator by time reversal. The direction of propagation has been reversed and gain & loss are interchanged but the delay is invariant. (c) The transformation of the oscillator by parity conjugation. $\mathbf{J}$ is the exchange matrix (the first Pauli matrix). The two loops are interchanged, and the coupler is reversed and inverted. (d) The transformation of the oscillator by parity conjugation and-time reversal. The original direction of propagation is restored.

# $\mathcal{PT}-$symmetric cross-injection dual optoelectronic oscillator

$\mathcal{PT}$ symmetric system concepts were introduced by Bender & Boettcher in 1998 [2]. The basic idea is to couple together a subsystem that gains energy from an external reservoir with an identical but time reversed subsystem that loses the same amount of energy to the exterior reservoir [3]. The overall system is then invariant to $\mathcal{PT}-$ symmetry transformations. Depending on the coupling strength there is a phase transition between unbroken $\mathcal{PT}-$ symmetric and broken $\mathcal{PT}-$symmetric solutions of the system. A realisation of such a system is a cross-coupled dual oscillator with one oscillator operating above and the other operating equally below the gain threshold for oscillation. In practice, additional gain is provided so both oscillators initially operate above threshold. The additional gain is saturable and reduces towards unity as the oscillation becomes established. In the frequency domain the modes occur as neutral doublets in the unbroken $\mathcal{PT}-$ symmetry phase and as a degenerate stable (decaying) and unstable (growing) pair in the broken $\mathcal{PT}-$ symmetry phase. The expectation is that the phase transition will favour the growth of one mode over all others significantly hastening an initial winner-takes-all gain competition leading ultimately to a single-mode established oscillation. It is natural that there is a growing research literature on the topic of $\mathcal{PT}-$ symmetric OEOs [4-21] given that $\mathcal{PT}-$ symmetric laser research has been particularly fruitful [22] and OEOs are related to lasers as TDOs.

Generally, authors of papers on $\mathcal{PT}-$ symmetric OEOs first present a theory of a coupled pair of ring resonators (Figure 1) as a model system illustrating the unbroken to broken $\mathcal{PT}-$ symmetry phase transition, followed by a description of the specific OEO circuit architecture investigated, and finally experimental observations of the spectrum of the established oscillation before and after some adjustment to a variable component of the architecture showing a dominant single mode oscillation emerging from a multimode oscillation. The phase noise performance, if reported, is generally poor, circa $-100\ dBc/Hz\ @\ 10\ kHz$ offset from a 10 GHz carrier. The evidence that the theoretical model is an accurate description of the proposed circuit architecture is tenuous. No simulations of the circuit architecture are performed and the experimental observation of the emergence of single mode oscillation following a manual adjustment of oscillator components is not conclusive evidence of a $\mathcal{PT}-$ symmetry transition since it can also be the outcome of a gain competition process.

The model system ring resonators are nominally identical except that one has gain and the other loss. The interaction between the rings is modelled as a time variation of the complex amplitude of a natural mode common to each ring [23]. The result is a multitude of coupled pairs of first order *linear* differential equations indexed by the common mode index. This model describes the dynamics by a complex amplitude that depends on time only, i.e., it is assumed that the coupling is sufficiently weak that variation of the amplitude over one round-trip time is negligible. The validity of this assumption is dubious given the long delay ($\sim 25\ \mu s$) and the superposition of modes of comparable magnitude by the splitter/combiner components. It is also assumed that of the multitude of mode pairs, only one will experience the phase transition, i.e., the phase transition depends strongly on the natural frequency of the mode. Although, the formulation involves potentially distinct gain and coupling parameters indexed by the mode index, it is not clear how these parameters are determined. Contrary to the emphasis by the authors on wideband operation to the extent of removing the high-Q resonator of an orthodox OEO, a gain profile is sometimes invoked. A gain profile is normally understood to mean the dispersion of the common part of the gain, but the phase transition frequency is determined by the dispersion of the differential part of the gain and the coupling parameter, which are set by broadband passive components. Consequently, the phase transition is global in its effect on the modes.

## 𝒫𝒯 −symmetric cross-injection dual optoelectronic oscillator

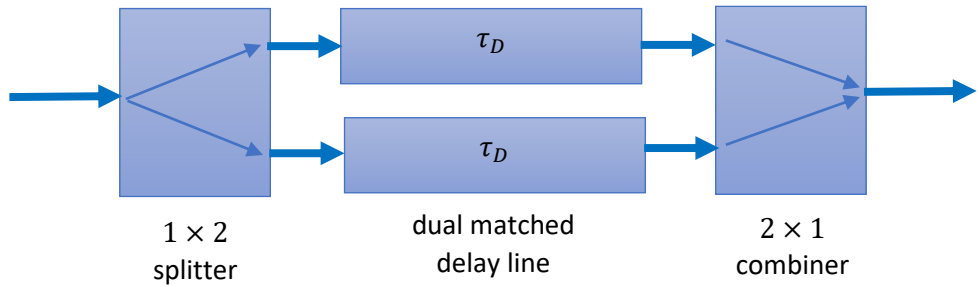

(a) Dual matched RF-photonic loop with dimension reduction. Typically, the splitter is implemented in the optical domain following intensity modulation by the RF input and the combiner is implemented in the electronic domain after photodetection.

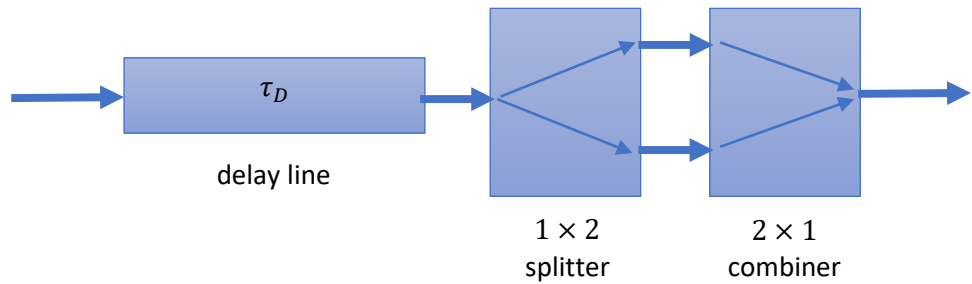

(b) Equivalent *single* loop equivalent to the dual matched loop in (a)

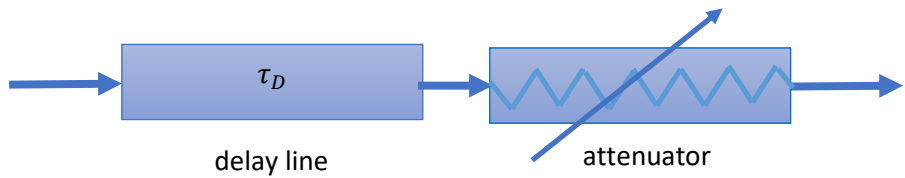

(c) Redrawing of (b) with the back-to-back splitter / combiner replaced by an attenuator which may be variable depends on the implementation of the splitter/combiner and their transmission parameters.

*Figure 2. An Illustration of the equivalence of a dual matched delay line with delay time $\tau_D$ placed between a $1 \times 2$ and a $2 \times 1$ coupler to a single delay line and attenuator. The attenuation depends on the transmission of the splitter/combiner. The attenuation may be compensated by the RF amplifier. Manually adjusted, the attenuation will have an influence on the gain competition process, e.g., a transient reduction of gain can extinguish modes close to the oscillation threshold favouring an above-oscillation threshold mode in the ensuing competition for the increased gain.*

Moreover, the theory can only describe the brief initial linear regime of oscillation. $\mathcal{PT}$ − symmetry related phenomena in the linear regime have been experimentally investigated using an electronic oscillator with two coupled LCR resonators [24] but successful measurements required an exceptionally low natural frequency $\sim 30 \, kHz$ and great experimental care. In practice, measurements on an OEO are limited to established oscillations in the saturation regime. Consequently, an extension of the theory to the saturation regime is a prerequisite to the interpretation of experimental measurements.

The circuit architectures presented most often have two RF-photonic paths that are closed into a loop by a *single* shared RF-electronic path. An exception is reference [5] where two RF-electronic paths are used. The RF-photonic path has two RF inputs that intensity modulate optical carriers that are polarisation multiplexed onto a single optical fibre. However, the necessity of demultiplexing the modulated carriers is overlooked; a polarisation insensitive optical coupler produces two copies of the same optical intensity on the two photodiodes, thereby forming the same linear superposition of the RF signals at the two outputs. Consequently, in all cases in the prior art, the two-dimensional RF vector signal is reduced to a one-dimensional RF scalar signal within some segment of the oscillator loop. Significant effort has been expended in matching the delay of the two RF-photonic paths and ensuring robustness to environmental fluctuations, e.g., by using two optical carriers of distinct frequency to carry as modulation two distinct RF signals [7] but the necessity of demultiplexing is systematically overlooked. The consequences are profound. Figure 2 illustrates that a dual matched delay line placed between a $1 \times 2$ and a $2 \times 1$ coupler is equivalent to a single delay line with the same delay. The dual loop OEO thereby degenerates into a single-loop OEO. The dimension reduction causes the transmission matrix of the coupler between the two rings to have zero determinant which results in a globally broken $\mathcal{PT}$ − symmetry phase (see Equation 15 with $\det(\mathbf{K}) = 0$). Mode selection can only occur through the gain competition process which is rendered ineffective by the absence of a high-Q RF resonator. It is conjectured that a transient reduction in gain caused by intervention of the experimentalist ostensibly to adjust the gain/loss parameter is responsible for the reported observation of mode selection.

This paper carefully addresses each of the pitfalls encountered in the prior art. A delay differential / integral equation is used to model a $\mathcal{PT}$ − symmetric oscillator. Scattering matrix methods are used to construct the $\mathcal{PT}$ − invariant transmission matrix characterising the coupler. The coupler transmission matrix is necessarily invertible, and no assumption of weak coupling is made. A first decomposition of the coupler transmission matrix enables the identification of a circuit architecture that can be implemented using standard RF-photonic components except for the sustaining amplifier which requires a particular saturation mechanism to preserve $\mathcal{PT}$ − invariance. A mode-selective $\mathcal{PT}$ − symmetry phase transition is achieved by placing a short delay within one arm of an MZI that appears within the circuit architecture. A second decomposition of the coupler transmission matrix enables the identification of a differently configured but equivalent circuit architecture corresponding to a cross-injection dual TDO with similar but distinct delays. The localised $\mathcal{PT}$ − symmetry phase transition may then be understood in terms of the Vernier effect. The cross-injection perspective facilitates the extension of the theory to a phase-only model of an established oscillation that inter alia provides an analytic prediction of the phase-noise spectral density. Advantageously, the second configuration enables the special sustaining amplifier to be replaced by pair of matched but otherwise independent amplifiers. Thereby, the second configuration is amenable to practical implementation as a dual OEO using standard RF-photonic and RF-electronic components throughout. Both configurations of the circuit architecture map neatly and expressively into a Simulink™ simulation model. Extensive simulation trials have been performed to validate the theory, observe known $\mathcal{PT}$ − symmetry breaking phenomena, confirm the equivalence of the two configurations, and reproduce the phase-noise spectral density predicted by the analytic theory. There is excellent agreement between the theoretical and simulation results.

Prior investigations of cross injection dual OEOs have been reported in the literature. In a pioneering paper [25]. a comparison of the predictions of a theoretical model with experimental measurements is presented. However, recourse to a comprehensive computational model described in [26] is necessary to obtain a reasonable fit to experimental results. Moreover, consistent with a master-slave oscillator perspective, disparate delays of $20\ \mu s$ (master) and $0.22 - 2.74\ \mu s$ (slave) are implemented rather than 𝒫𝒯 −symmetric cross-injection dual optoelectronic oscillator# 𝒫𝒯 −symmetric cross-injection dual optoelectronic oscillator the similar delays required for clear observations of a mode-selective 𝒫𝒯 − symmetry phase transition and the related Vernier effect.

The paper is structured as follows. Section 2 introduces the 𝒫𝒯 − symmetric circuit architecture, derives the properties of the coupler transmission matrix and its eigenvalue structure, describes their impact on the oscillator behaviour, and proposes a mechanism for a mode-selective 𝒫𝒯 − symmetry phase transition. The equivalence to a symmetric cross-injected dual time delay oscillator is established and an injection locking and phase noise analysis provided. Section 3 introduces the Simulink™ simulation models, explains their principal features, and provides representative simulation results. The paper concludes with Section 4 which provides a summary of the main findings and a discussion of their significance. Appendix I provides details of the scattering matrix method and Appendix II provides details of how the injection locking theory of a time delay oscillator introduced in reference [27] is extended to a two-dimensional oscillator.

## 2. Theory

### 2.1.   𝒫𝒯 − symmetric circuit architecture

Figure 1 provides a schematic illustration of a coupled dual TDO composed of a pair of identical rings with delay time $\tau_D$ coupled by a linear time invariant (LTI) subsystem represented by the operator 𝒦 characterised by its transmission matrix **K**. The oscillator may be modelled by the two-dimensional system of delay differential / integral equations:

$$\boldsymbol{u}(t) = \kappa \boldsymbol{\mathcal{K}} \boldsymbol{u}(t - \tau_D)$$

*Equation 1*

where $\boldsymbol{u}$ is the oscillator state-vector with components equal to the complex envelope of the oscillation at the exit ports of the coupler system, and $\kappa$ is a scalar valued overall gain parameter which may absorb any overall phase factor, excess gain or excess loss contributed by other components.

The subsystem described by $\kappa\boldsymbol{\mathcal{K}}$ includes a dual RF amplifier that sustains the oscillation and a dual matched bandpass filter providing an overall gain profile. The dual RF amplifier is treated as the composition of a dual matched saturating amplifier and a pair of linear amplifiers providing matched gain and loss. The dual RF amplifier is placed between a pair of MZIs which acts as variable lossless couplers controlled by the differential phase shift of the interferometer arms (see Figure 1, Figure 6, & Figure 7).

The saturated gain is the describing function [28] of the harmonic balance method [29] (a special case of the Galerkin method [30]) applied to provide a quasi-linear description of the nonlinearity of a saturating amplifier. It is fundamental to the method that the harmonics generated by the nonlinearity are ultimately dissipated by the bandpass filtering action of the remaining components in the oscillator loop. The method captures the amplifier saturation as a gain parameter that is a function of the magnitude of the oscillation at the amplifier input port. In general, the magnitudes of the oscillation at the dual matched saturating amplifier input ports differ and, while the linear gain of each amplifier agrees, their saturated gain does not. To rectify this shortcoming, rather than the saturated gain of each amplifier being a function of the magnitude $|u_1|, |u_2|$ of the oscillation at its respective input port, it is treated as a function of the norm $\|\boldsymbol{u}\| = \sqrt{|u_1|^2 + |u_1|^2}$ of the vector of inputs to the dual amplifier. Thereby the saturated gain of both amplifiers remains matched in the saturation regime advantageously rendering the dual matched saturating amplifier invariant to unitary transformations $U(2)$.

66

# $\mathcal{PT}$ −symmetric cross-injection dual optoelectronic oscillator

The transmission matrix describing the $U(2)-$ invariant dual matched saturating amplifier block is the product of the scalar-valued gain control function $\kappa$ and the identity. The gain control function is the only nonlinearity considered in the theory; the remainder of the oscillator model is treated as linear. The $U(2)-$ invariant dual matched saturating amplifier block commutes with all other blocks within the oscillator loop except for the matched linear gain/loss block. Nevertheless, it can be placed either side of the matched gain/loss block by taking the long route around the oscillator loop. Hence, the choice of location within the oscillator loop of the $U(2)-$invariant dual matched saturating amplifier is free.

In the case of an OEO, each delay line is implemented as an RF photonic link consisting of a laser followed by a Mach-Zehnder modulator (MZM) driven by the RF input that provides an optical intensity modulated input to an optical fibre delay line which is terminated by a photo-receiver that provides the RF output. The MZM has a sinusoidal optical intensity transmission versus RF voltage characteristic, which is the principal source of non-linearity of the RF delay line. Prima facie, a linear treatment either restricts analysis to the initial linear oscillator regime in which the RF amplitude is small or requires a means of linearizing the modulator transfer function in the case of a fully established oscillation. The issue is not fundamental as there are variety of means of linearizing an MZM [31].

An optical amplifier is an example of a physical system possessing a saturation mechanism dependent on the sum of input powers, equivalently, the squared norm $\|\boldsymbol{u}\|^2$. This is the basis of cross-gain modulation using a polarisation insensitive optical amplifier or a two-wavelength optical amplifier where an in-homogeneously broadened gain medium with a common pumping mechanism supports two distinguishable probe signals. In the case of an electronic amplifier, it is a challenge to match the inherent speed of gain control by a limiting amplifier. An automatic gain control circuit (AGC) might be effective that has sufficient speed to suppress amplitude fluctuations at frequencies within the passband of the bandpass filter (BPF) while leaving the BPF to dissipate the higher frequency fluctuations. This expedient might profit from the reduced flicker noise of linear amplifiers compared to amplifiers operating in saturation. It turns out, however, that a judicious choice of location within the oscillator loop of the $U(2)-$ invariant dual matched saturating amplifier permits the substitution of a matched pair of independent limiting amplifiers without disturbance to the oscillator behaviour. Moreover, this expedient also renders linearization of the MZMs unnecessary as their nonlinear transfer function may be subsumed within the saturation behaviour of the limiting amplifiers.

The LTI operator $\boldsymbol{\mathcal{K}}$ is a $2 \times 2$ matrix of convolutional operators fully characterised *in the Fourier domain* by a $2 \times 2$ complex matrix valued transmission function $\mathbf{K}$. The passband of the BPF present within each loop is broad and has negligible effect on slowly varying phenomenon. For initial considerations the variable couplers are considered adjustable but not frequency dependent. The assumption of broadband operation is equivalent to $\mathbf{K}(\omega) \sim \mathbf{K}(0)$ where $\omega$ is an *offset frequency* from the nominal operating frequency. The zero offset frequency is dropped in the sequel for simplicity of notation. The constant transmission matrix may then be pulled out of the Fourier transform and directly applied in the time domain, so Equation 1 simplifies to:

$$\boldsymbol{u}(t) = \kappa \mathbf{K} \boldsymbol{u}(t - \tau_D)$$

*Equation 2*

The transmission function may be normalised as convenient to the problem by suitably scaling the gain control function $\kappa$ so that the product $\kappa \mathbf{K}$ remains invariant. The broadband assumption is relaxed in the sequel.

## 2.2. $\mathcal{PT}$ − invariance

Appendix I introduces the action on an arbitrary transmission matrix **T** of the operation of parity conjugation $\mathcal{P}$ and time reversal $\mathcal{T}$:

$$\mathbf{T} \xrightarrow{\mathcal{P}} \mathbf{J}_1 \mathbf{T}^T \mathbf{J}_1 \quad ; \quad \mathbf{T} \xrightarrow{\mathcal{T}} (\mathbf{T}^{-1})^* \quad ; \quad \mathbf{J}_1 = \begin{bmatrix} 0 & 1 \\ 1 & 0 \end{bmatrix}$$

*Equation 3*

where the exchange matrix in two dimensions is represented by the first Pauli matrix $\mathbf{J}_1$.

The transmission matrix of a lossless dual matched delay line:

$$\mathbf{D}_{\tau_D} = \exp(-i\omega\tau_D)\,\mathbf{I}$$

*Equation 4*

is $\mathcal{P}$ −invariant and $\mathcal{T}$ −invariant. Consequently, the $\mathcal{PT}$ −symmetry properties of the oscillator are determined by the LTI system which is $\mathcal{PT}$ − invariant only if its transmission matrix **K** satisfies:

$$\mathbf{K} = \mathbf{J}_1 (\mathbf{K}^{-1})^\dagger \mathbf{J}_1$$

*Equation 5*

A straightforward computation shows that the transmission matrix is of the form:

$$\mathbf{K} = \begin{bmatrix} a & ib \\ ic & d \end{bmatrix} \quad ; \quad \det(\mathbf{K}) = 1 \quad ; \quad a, b, c, d \in \mathbb{R}$$

*Equation 6*

It is shown in Appendix I that a $\mathcal{PT}$ −invariant transmission matrix admits the decomposition:

$$\mathbf{K} = \mathbf{U}(\beta)\boldsymbol{\Gamma}(\gamma)\mathbf{U}(\alpha) \quad ; \quad \alpha, \beta, \gamma \in \mathbb{R}$$

*Equation 7*

where:

$$\mathbf{U}(\theta) = \begin{bmatrix} \cos(\theta) & i\sin(\theta) \\ i\sin(\theta) & \cos(\theta) \end{bmatrix}$$

*Equation 8*

is the transmission matrix of a variable coupler and:

$$\boldsymbol{\Gamma}(\gamma) = \begin{bmatrix} \exp(\gamma) & 0 \\ 0 & \exp(-\gamma) \end{bmatrix}$$

*Equation 9*

is the transmission function of the matched linear gain/ loss block.

The variable coupler transmission matrix $\mathbf{U}(\theta)$ admits the decomposition:

$$\mathbf{U}(\theta) = \mathbf{H}\exp(i\theta\mathbf{J}_3)\,\mathbf{H} \quad ; \quad \mathbf{H} = \frac{1}{\sqrt{2}}\begin{bmatrix} 1 & 1 \\ 1 & -1 \end{bmatrix} \quad ; \quad \mathbf{J}_3 = \begin{bmatrix} 1 & 0 \\ 0 & -1 \end{bmatrix}$$

*Equation 10*

which is recognised as describing the transmission of an MZI with a differential phase shift of $\theta$ in its arms represented by $\exp(i\theta\mathbf{J}_3)$ formed between two 180° hybrid couplers represented by **H**; a real Hadamard matrix in Sylvester's form normalised to be orthogonal. $\mathbf{J}_3$ is the third Pauli matrix.

# $\mathcal{PT}-$symmetric cross-injection dual optoelectronic oscillator

### 2.2.1. Eigenvalue structure

The substitutions:

$$\boldsymbol{u} = \boldsymbol{a}\exp(st) \quad ; \quad s = \sigma + i\omega$$

*Equation 11*

$$\kappa \exp(-s\tau_D)\lambda = 1$$

*Equation 12*

transform Equation 2 into the algebraic eigenvalue-eigenvector problem:

$$\mathbf{K}\boldsymbol{a} = \lambda \boldsymbol{a}$$

*Equation 13*

which has a characteristic polynomial:

$$\lambda^2 - \text{tr}(\mathbf{K})\lambda + \det(\mathbf{K}) = 0$$

*Equation 14*

Equation 12 is an expression of the Barkhausen oscillation condition.

The characteristic equation has *real* coefficients on account of Equation 6. Since the trace is equal to the sum of the eigenvalues and the determinant is equal to their product, the eigenvalues either form a complex conjugate pair corresponding to solutions with *unbroken $\mathcal{PT}-$symmetry* or a reciprocal pair on the real axis corresponding to solutions with *broken $\mathcal{PT}-$symmetry*.

Explicitly, the eigenvalues are given by:

$$\lambda_\pm = \text{tr}(\mathbf{K})/2 \pm \sqrt{(\text{tr}(\mathbf{K})/2)^2 - \det(\mathbf{K})}$$

*Equation 15*

$$\det(\mathbf{K}) = 1$$

*Equation 16*

$$\text{tr}(\mathbf{K})/2 = \cosh(\gamma)\cos(\alpha + \beta)$$

*Equation 17*

where use has been made of the one parameter subgroup property:

$$\mathbf{U}(\theta_1 + \theta_2) = \mathbf{U}(\theta_1)\mathbf{U}(\theta_2)$$

*Equation 18*

and the invariance of the trace to cyclic permutations of a matrix product.

In the context of an oscillator, Equation 17 indicates that one of the variable couplers is redundant and may be removed.

#### Unbroken $\mathcal{PT}-$ symmetry regime

The discriminant in Equation 15 is negative for:

$$|\cosh(\gamma)\cos(\alpha + \beta)| < 1$$

*Equation 19*

which yields a complex conjugate pair of eigenvalues on the unit circle:

$$\lambda_\pm = \exp(\pm i \nu \tau_D)$$

*Equation 20*

$$\nu\tau_D = \tan^{-1}\left(\frac{\sqrt{1 - \cosh^2(\gamma)\cos^2(\alpha + \beta)}}{\cosh(\gamma)\cos(\alpha + \beta)}\right)$$

*Equation 21*

# $\mathcal{PT}-$symmetric cross-injection dual optoelectronic oscillator

The arctangent in Equation 21 may be interpreted in the four-quadrant sense to account for a change in sign of the trace and may be unwrapped if the parameters are continuous functions of a variable such as frequency.

The Barkhausen oscillation condition takes the form:

$$\omega\tau_D = \arg(\kappa) + 2p\pi \pm \nu\tau_D \quad p \in \mathbb{Z}$$
$$\sigma\tau_D = \ln(|\kappa|)$$

*Equation 22*

For $|\kappa| = 1$ the modes are neutral, i.e., they neither grow nor decay with time.

The resonances occur in doublets with each resonance of the doublet displaced in frequency by $\pm\nu$ from the nodes of a regular grid with interval $2\pi/\tau_D$. The maximum doublet splitting is $\pi/\tau_D$ corresponding to a regular grid with interval $\pi/\tau_D$. An alternative perspective is that the resonances correspond to two interlaced regular frequency combs with interval $2\pi/\tau_D$ that are displaced respectively by $\pm\nu$ from complete alignment. In the absence of gain/loss $\gamma = 0$ and Equation 21 reduces to:

$$\nu\tau_D = \alpha + \beta$$

*Equation 23*

illustrating the role played by $\alpha, \beta$ in tuning the displacement between the two frequency combs.

*Broken $\mathcal{PT}-$ symmetry regime*

The discriminant in Equation 15 is positive for:

$$|\cosh(\gamma)\cos(\alpha + \beta)| > 1$$

*Equation 24*

which for positive $\cos(\alpha + \beta)$ leads to a reciprocal pair of eigenvalues on the positive real axis:

$$\lambda_{\pm} = \exp(\pm\mu\tau_D) \quad ; \quad \mu\tau_D = \ln\left(\frac{\sqrt{\cosh^2(\gamma)\cos^2(\alpha + \beta) - 1}}{|\cosh(\gamma)\cos(\alpha + \beta)|}\right)$$

*Equation 25*

corresponding to a degenerate pair of modes at each even integer node of a regular frequency grid with interval $\pi/\tau_D$:

$$\left.\begin{array}{l}\sigma = \ln(|\kappa|) \pm \mu\tau_D \\ \omega\tau = \arg(\kappa) + 2p\pi\end{array}\right\} \quad p \in \mathbb{Z}$$

*Equation 26*

or for negative $\cos(\alpha + \beta)$ leads to a reciprocal pair of negative eigenvalues on the negative real axis:

$$\lambda_{\pm} = -\exp(\pm\mu\tau_D)$$

*Equation 27*

corresponding to a degenerate pair of modes at each odd integer node of a regular frequency grid with interval $\pi/\tau_D$:

$$\left.\begin{array}{l}\sigma = \ln(|\kappa|) \pm \mu\tau_D \\ \omega\tau = \arg(\kappa) + (2p+1)\pi\end{array}\right\} \quad p \in \mathbb{Z}$$

*Equation 28*

For $|\kappa| = 1$ one mode is stable (grows with time) and the other is unstable (decays with time).

# $\mathcal{PT}-$symmetric cross-injection dual optoelectronic oscillator

*The exception point*

The unbroken and broken symmetry regimes are delineated by a zero discriminant that occurs for:

$$\cosh^2(\gamma)\cos^2(\alpha+\beta) = 1$$

*Equation 29*

The combination of parameters satisfying this equation is known as the exception point. As this point is approached the two eigenvalues approach each other to form in the limit a unit magnitude double real root, i.e., the algebraic multiplicity is 2. The two eigenvectors associated with the two eigenvalues also approach each other and coincide in the limit up to normalisation. There is only one eigenvector at the exception point, i.e., the geometrical multiplicity is 1. Consequently, at the exception point the matrix **K** does not possess a complete basis of eigenvectors and cannot be diagonalized. The matrix is said to be defective.

## 2.3. Mode-selective $\mathcal{PT}$-symmetry phase transition

The BPF plays an important role in promoting single-mode oscillation of a conventional TDO. Initially the sustaining amplifier operates in its linear regime and oscillation builds up from an initial transient or fluctuation. A potential oscillation mode grows at a rate proportional to the net gain it experiences on each round trip. A suitably tuned BPF with a bell-shaped transmission function magnitude favours the growth of one mode above all others. The gain control mechanism, usually saturation of the sustaining amplifier, introduces a winner-takes-all competition by reducing the round-trip gain as the oscillation magnitude grows until the favoured mode is sustained by unit net gain and all other modes experience net loss and decay. However, a BPF with a wide passband is advantageous for broadband operation and a long delay is advantageous for low phase noise. A wide passband and long delay render mode selection challenging because a very large number of potential oscillating modes fall within the passband of the BPF and the desired mode then receives little advantage compared its neighbours. A $\mathcal{PT}-$ symmetric TDO provides a mechanism that accelerates gain competition via the increased gain contrast between modes with broken $\mathcal{PT}-$ symmetry and modes with unbroken $\mathcal{PT}-$ symmetry.

However, if the $\mathcal{PT}$-symmetry phase transition is induced by manually adjusting the value of the otherwise constant parameters $\alpha+\beta$, $\gamma$, the eigenvalue analysis shows that the $\mathcal{PT}$-symmetry phase transition is *global*. Although half the modes experience an increase in gain and the remainder experience increased loss, the gain contrast among the modes with increased gain is unchanged. The node selection proceeds via the same mode competition mechanism as a single oscillator with the same increment in overall gain. The gain profile favours one mode above the other surviving modes no more after than before the phase transition.

To appreciate this point more fully, consider the special case $\alpha = \beta = 0$ which implies $\mathbf{K} = \mathbf{\Gamma}(\gamma)$. The dual oscillator thereby decomposes into two independent TDOs, one with a loop containing the gain and one with a loop containing the loss. For $|\kappa|=1$ the oscillator containing the loss cannot oscillate. Only the oscillator containing the gain oscillates and mode selection occurs via the same process of mode competition as a standard single loop TDO. The same outcome is assured by the $U(2)-$ invariant dual matched saturating amplifier even if its linear gain is sufficient for both oscillators to initially to oscillate. The oscillator with loop containing the gain will rapidly suppress oscillation of the oscillator with the loop containing the loss term as the gain of the dual amplifier saturates. Moreover, for the general case of $\cos^2(\alpha+\beta)=1$, the decomposition into independent TDOs remains valid albeit the loops are spatially multiplexed within the shared dual matched delay lines.

# $\mathcal{PT}$−symmetric cross-injection dual optoelectronic oscillator

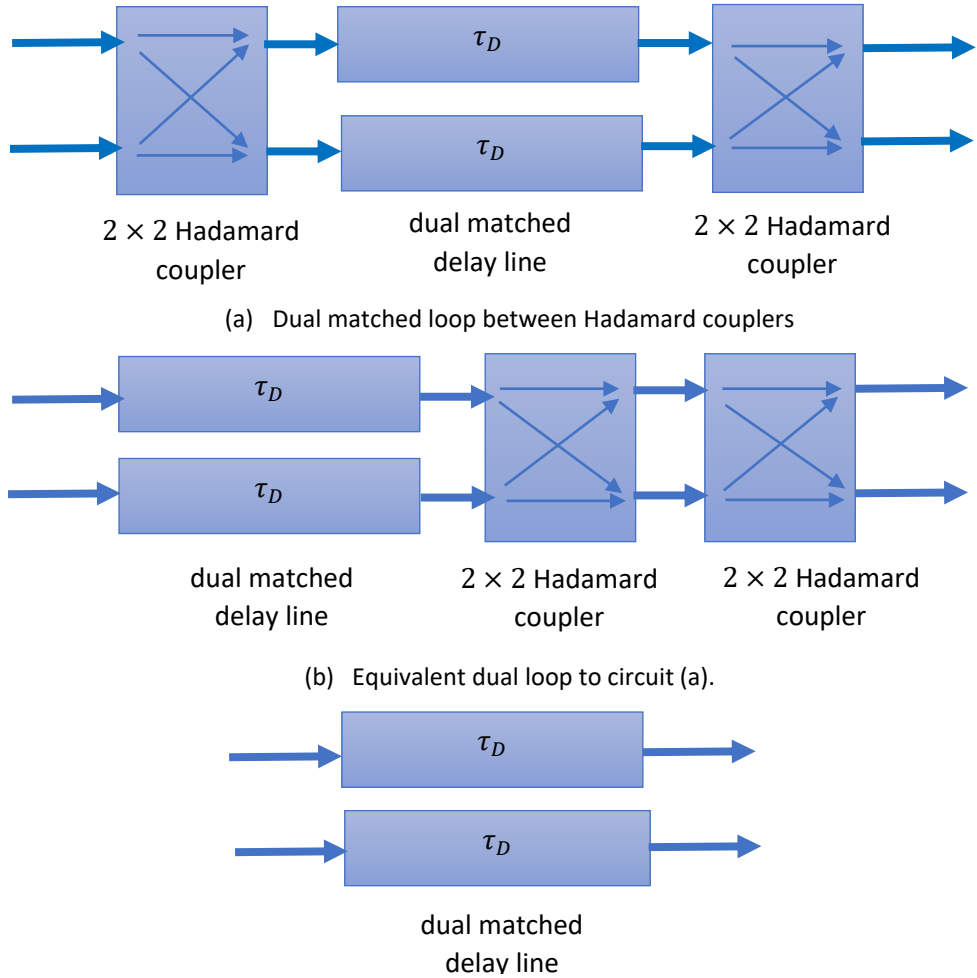

(a) Dual matched loop between Hadamard couplers

(b) Equivalent dual loop to circuit (a).

(c) The Hadamard couplers in (b) annihilate leaving a dual matched delay line without couplers.

Figure 3. An illustration of the invariance of a dual matched delay-line with delay time $\tau_D$ to conjugation by a $2 \times 2$ Hadamard coupler (a 180°- hybrid coupler).

The $\mathcal{PT}$-symmetry phase transition must be *local* to be mode-selective, i.e., the phase transition must depend on frequency. In $\mathcal{PT}-$ symmetric lasers a frequency dependent $\gamma$ has been achieved using an alternating gain and loss grating [32]. In the case of a $\mathcal{PT}-$ symmetric TDO, it is more convenient to engineer a frequency dependent $\alpha$, $\beta$ using MZI arms with differential delay $\tau_\alpha$, $\tau_\beta$ so that:

$$\alpha = \alpha_0 + \omega\tau_\alpha \quad ; \quad \beta = \beta_0 + \omega\tau_\beta$$

*Equation 30*

where $\alpha_0$ & $\beta_0$ are static phase biases. This can be achieved by inserting a short delay line in one arm only of each interferometer with a delay time equal to $2\tau_\alpha, 2\tau_\beta$. The common part of the delay introduced within the two arms required to preserve causality adds to the delay provided by the dual matched delay line. The frequency-dependent phase transition thereby introduced is periodic in frequency, so an overall gain profile remains necessary to suppress the undesired periodic revival of modes with broken $\mathcal{PT}-$ symmetry. There is no entire dispensing with the BPF, but the requirement on its passband width is relaxed in a $\mathcal{PT}-$ symmetric compared to a standard TDO.

## 2.4. $SU(1,1) -$ symmetric configuration

In Appendix I it is shown that a $\mathcal{PT}-$symmetric coupler transmission matrix admits the decomposition:

$$\mathbf{K} = \mathbf{HVH}$$

*Equation 31*

where $\mathbf{V} \in SU(1,1)$.

It follows from $\mathbf{H}^{-1} = \mathbf{H}$, that $\mathbf{V}$ and $\mathbf{K}$ are *similar* matrices and consequently share the same eigenvalues. Moreover, the behaviour of the oscillator is unchanged by the removal of the Hadamard matrices in the decomposition of Equation 31 since they commute with the dual matched delay line and thereby annihilate as illustrated in Figure 3. The eigenvalues are invariant to conjugation by the Hadamard matrices, only the eigenvectors change. Similarity is an *equivalence relation*[1].

The transmission matrix $\mathbf{V}$ admits the representation:

$$\mathbf{V} = \exp(i\beta \mathbf{J}_3) \exp(\gamma \mathbf{J}_1) \exp(i\alpha \mathbf{J}_3)$$

*Equation 32*

With the removal of the outer pair of two 180° hybrid couplers represented by $\mathbf{H}$, the remainder of the $\mathcal{PT}-$symmetric oscillator loop consists of the dual matched delay line represented by the diagonal transmission matrix $\mathbf{D}_{\tau_D}$ defined by Equation 4. The arms of the former MZIs represented by $\exp(i\alpha \mathbf{J}_3)$ and $\exp(i\beta \mathbf{J}_3)$ then directly concatenate with the dual matched delay line permitting the consolidation of the component delay lines and phase shifters to form a pair of *independent delay lines with distinct delay times and tuning phase-shifts*. This equivalence is illustrated in Figure 4.

After the consolidation, all that remains of $\mathbf{V}$ is the second term of the decomposition; a real symmetric matrix:

$$\exp(\gamma \mathbf{J}_1) = \mathbf{H}\Gamma(\gamma)\mathbf{H} = \begin{bmatrix} \cosh(\gamma) & \sinh(\gamma) \\ \sinh(\gamma) & \cosh(\gamma) \end{bmatrix}$$

*Equation 33*

belonging to a one parameter subgroup of $SU(1,1)$. This matrix may be alternatively written:

$$\exp(\gamma \mathbf{J}_1) = \cosh(\gamma) \, \mathbf{P} \quad ; \quad \mathbf{P} = \begin{bmatrix} 1 & \rho \\ \rho & 1 \end{bmatrix} \quad ; \quad \rho = \tanh(\gamma)$$

*Equation 34*

and the factor $\cosh(\gamma)$ absorbed into the scalar gain control parameter $\kappa$ yielding a symmetric cross-injection coupler transmission matrix:

$$\mathbf{P} = \begin{bmatrix} 1 & \rho \\ \rho & 1 \end{bmatrix} \quad ; \quad \rho = \tanh(\gamma)$$

*Equation 35*

realisable using passive components. The parameter $\rho$ is recognised as the cross-injection ratio. Since $\mathbf{P}$ is real and bisymmetric, best performance is attained if splitting ratio, loss, and the relative phase of the through- and cross- paths are matched.

---

[1] A binary relation $\sim$ is an equivalence relation if and only if it is reflexive $a{\sim}a$, symmetric $a{\sim}b \implies b{\sim}a$, and transitive $a{\sim}b \, \& \, b{\sim}c \implies a{\sim}c$.

# $\mathcal{PT}-$symmetric cross-injection dual optoelectronic oscillator

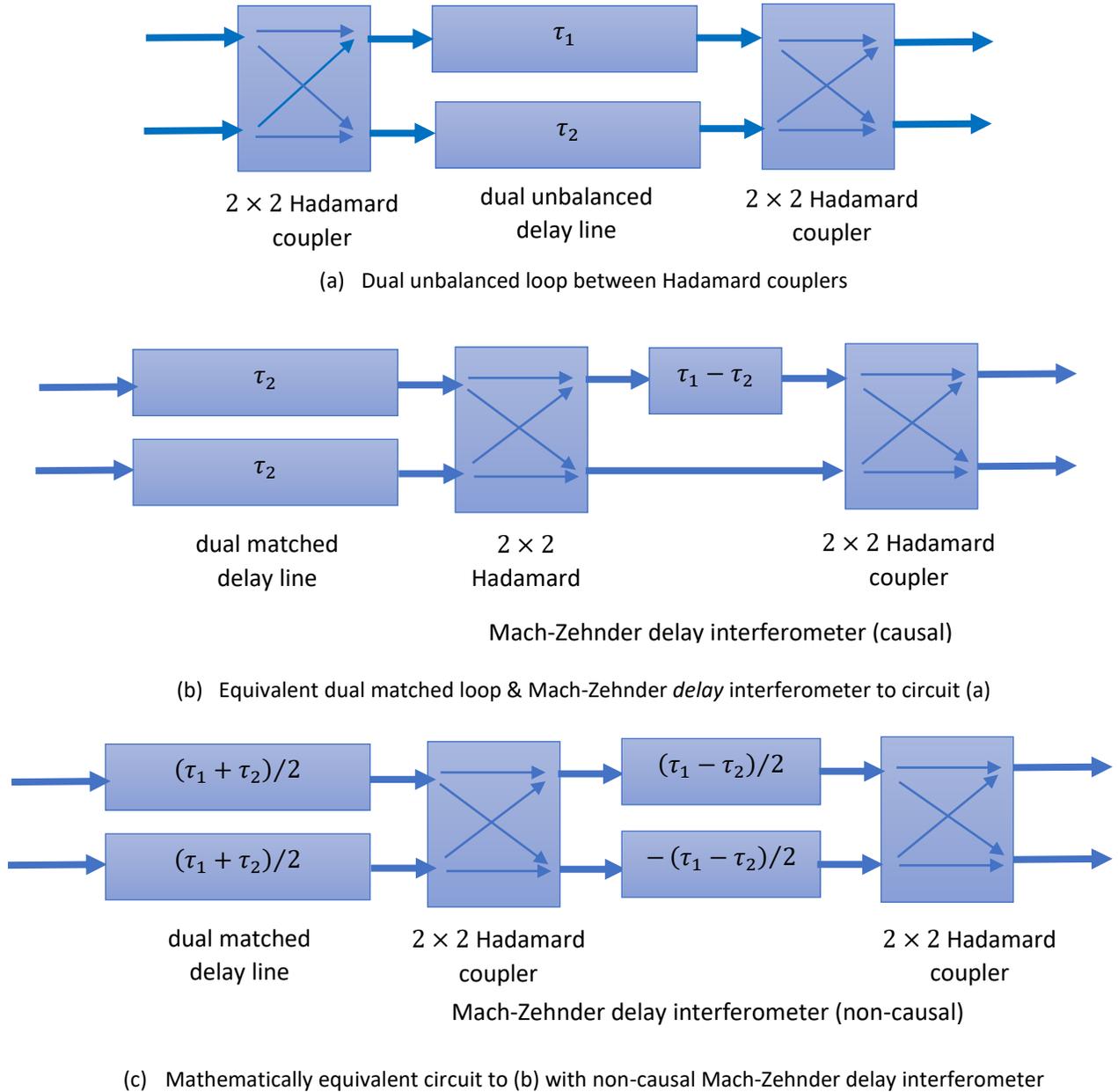

(a) Dual unbalanced loop between Hadamard couplers

(b) Equivalent dual matched loop & Mach-Zehnder *delay* interferometer to circuit (a)

(c) Mathematically equivalent circuit to (b) with non-causal Mach-Zehnder delay interferometer

*Figure 4. An illustration of the equivalence of (a) an unbalanced dual delay line with delay times $\tau_1, \tau_2$ placed between two $2\times 2$ Hadamard couplers to (b) & (c) a dual matched delay concatenated with a Mach-Zehnder delay interferometer. A physical implementation must use causal components as in (b). For analysis, one is free to use noncausal components provided the complete subsystem is causal as in (c). The Mach-Zehnder delay interferometer in (c) has a transmission matrix that is the same form as a lossless variable coupler (see Equation 8 ) with a parameter $\theta = \omega(\tau_1 - \tau_2)/2$ that depends linearly on the oscillation frequency $\omega$.*

It has been established that a symmetric cross-injection dual TDO with unequal delay times is *equivalent* to a $\mathcal{PT}-$ symmetric dual TDO with mode selective $\mathcal{PT}-$ symmetry phase transition. The eigenvectors differ between the two configurations. Nevertheless, the $U(2)$ invariance of the dual matched amplifier ensures that it may be placed in any location within the oscillation loops of either configuration. The use of limiting amplifiers placed between the dual delay line and cross-injection coupler as the sustaining amplifiers of a cross-injection locked dual TDO presents no issues. Consequently, it is intuitive that, without disturbance to the oscillator, one may substitute a matched but otherwise *independent* pair of limiting amplifiers for a $U(2)$ invariant dual matched amplifier *placed at this specific location*. Certainly, in the linear regime the linear gains agree and in the saturated regime the saturated gains and output powers also agree.

For negligible detuning of the desired mode between the oscillators, the cross-injection locks the relative phase of the oscillators. Consequently, if the oscillation is transformed by a coupler with a Hadamard transmission matrix, the oscillator will establish a bright and dark port at the coupler output. This occurs at the gain/loss block in the $\mathcal{PT}-$ symmetric configuration as the second term of Equation 33 demonstrates. For zero detuning, matched RF amplifier saturation and ideal couplers, substantially complete suppression of the carrier at the dark port occurs. It is similarly intuitive that the oscillating mode in the $\mathcal{PT}-$ symmetric configuration avoids the path with loss through a process of self-organisation thereby creating a dark port. Consequently, in the $\mathcal{PT}-$ symmetric configuration the $U(2)$ invariance of a dual matched saturating amplifier located adjacent to the gain/loss block is vital to the correct operation of the oscillator in the saturated regime. It is the Hadamard conjugation equivalence relation combined with the self-organisation properties of the dual oscillator that enables standard limiting amplifiers to be substituted for the special $U(2)-$ invariant dual saturating amplifier. Compatibility with standard amplifier technology is a significant merit of the $SU(1,1)$ configuration.

## 2.5. Injection locking analysis.

The magnitude of an established oscillation is essentially held constant by the saturation of the sustaining amplifier so that amplitude fluctuations are substantially suppressed. It is consequently an excellent approximation in the saturation regime to reduce a complex envelope model to a phase-only model of an oscillator. While the $\mathcal{PT}-$ symmetric configuration and the $SU(1,1)-$ configuration are equivalent, it is convenient to select the $SU(1,1)-$ configuration for analysis as it is particularly amenable to an injection-locking theory interpretation. The circuit architecture may be considered to consist of an outer ring corresponding to the figure of eight loop containing two inner rings corresponding to the loops of the coupled component oscillators. Oscillation of the outer ring can occur if the inner rings are tuned so that a pair of resonances come into close alignment. Once established the oscillation of the outer ring quenches all the natural oscillations thereby locking the phase of the output of the component oscillators and suppressing their sidemodes. The driven component oscillators act as frequency comb phase filters with incommensurate free spectral ranges (frequency interval between resonances) thereby ensuring single mode oscillation with high sidemode suppression provided the periodic revival of aligned resonances due to the Vernier effect fall outside the passband of the BPFs.

### 2.5.1. Oscillator phase evolution

A phase-only model of a TDO under injection by an external source introduced in reference [27] is readily extended to a cross-injection dual TDO. Given the symmetry of the $SU(1,1)-$ configuration the coupled evolution equations of the extended model given in Appendix II reduce to:

$$\theta_1 = (h \otimes D_{\tau_1})\left[\phi_1 + \tan^{-1}\left(\frac{\rho \sin(\theta_2 - \theta_1)}{1 + \rho \cos(\theta_2 - \theta_1)}\right) + \theta_1\right]$$

*Equation 36*

$$\theta_2 = (h \otimes D_{\tau_2})\left[\phi_2 + \tan^{-1}\left(\frac{\rho \sin(\theta_1 - \theta_2)}{1 + \rho \cos(\theta_1 - \theta_2)}\right) + \theta_2\right]$$

*Equation 37*

# $\mathcal{PT}$ −symmetric cross-injection dual optoelectronic oscillator

where $\theta_1$, $\theta_2$ are the oscillation phases, $\tau_1$, $\tau_2$ are the delays, $\phi_1$, $\phi_2$ are the tuning phases of oscillator 1 and oscillator 2 respectively; $\rho$ is the cross-injection ratio, $h$ is the impulse response of the BPFs, and $D_\tau$ is a delay operator defined by:

$$(D_\tau u)(t) = u(t - \tau)$$

*Equation 38*

The convolution $h \otimes$ may be approximated by the identity operator for fluctuation frequencies well within the passband of the BPFs and assuming evolution of the free oscillators to a single mode state, in the absence of injection:

$$\theta_1(t) - \theta_1(0) = \omega_1 t \quad ; \quad \theta_2(t) - \theta_2(0) = \omega_2 t$$
$$\Rightarrow$$
$$\phi_1 = \omega_1 \tau_1 \quad ; \quad \phi_2 = \omega_2 \tau_2$$

*Equation 39*

Equation 36 and Equation 37 simplify to:

$$\theta_1(t) - \theta_1(t - \tau_1) = \left[\omega_1 \tau_1 + \tan^{-1}\left(\frac{\rho \sin(\theta_2 - \theta_1)}{1 + \rho \cos(\theta_2 - \theta_1)}\right) + \theta_1\right](t - \tau_1)$$

*Equation 40*

$$\theta_2(t) - \theta_2(t - \tau_2) = \left[\omega_2 \tau_2 + \tan^{-1}\left(\frac{\rho \sin(\theta_1 - \theta_2)}{1 + \rho \cos(\theta_1 - \theta_2)}\right) + \theta_2\right](t - \tau_2)$$

*Equation 41*

A steady locked state under cross injection is described by a solution of the form:

$$\theta_1(t) - \theta_1(0) = \theta_2(t) - \theta_2(0) = \omega_\infty t \quad ; \quad \theta_2(t) - \theta_1(t) = \theta_\infty$$

*Equation 42*

where $\omega_\infty$ and $\theta_\infty$ are respectively the asymptotic locked frequency and phase difference between the two oscillators. It follows that:

$$(\omega_\infty - \omega_1)\tau_1 = \tan^{-1}\left(\frac{\rho \sin(\theta_\infty)}{1 + \rho \cos(\theta_\infty)}\right)$$

*Equation 43*

$$(\omega_\infty - \omega_2)\tau_2 = -\tan^{-1}\left(\frac{\rho \sin(\theta_\infty)}{1 + \rho \cos(\theta_\infty)}\right)$$

*Equation 44*

Eliminating $\omega_\infty$ provides a necessary condition for a solution to exist:

$$(\omega_2 - \omega_1)\tau_0 = \tan^{-1}\left(\frac{\rho \sin(\theta_\infty)}{1 + \rho \cos(\theta_\infty)}\right) \quad ; \quad \frac{1}{\tau_0} = \frac{1}{\tau_1} + \frac{1}{\tau_2}$$

*Equation 45*

which yields the locking-range:

$$(\omega_2 - \omega_1)\tau_0 \in [-\sin^{-1}(\rho), \sin^{-1}(\rho)]$$

*Equation 46*

Eliminating the right-hand sides of Equation 43, Equation 44 yields an expression for the asymptotic oscillation frequency in the form of weighted average of the natural frequency of the individual free oscillators:

# $\mathcal{PT}-$symmetric cross-injection dual optoelectronic oscillator

$$\omega_\infty = \frac{\tau_1}{\tau_1 + \tau_2}\omega_1 + \frac{\tau_2}{\tau_1 + \tau_2}\omega_2$$

*Equation 47*

An application of the mean value theorem enables the coupled time difference equations Equation 40, Equation 41 to be approximated by coupled differential equations valid for slowly varying $\theta_1, \theta_2$:

$$\tau_1 \frac{d\theta_1}{dt} = \omega_1 \tau_1 + \tan^{-1}\left(\frac{\rho \sin(\theta_2 - \theta_1)}{1 + \rho \cos(\theta_2 - \theta_1)}\right)$$

*Equation 48*

$$\tau_2 \frac{d\theta_2}{dt} = \omega_2 \tau_2 - \tan^{-1}\left(\frac{\rho \sin(\theta_2 - \theta_1)}{1 + \rho \cos(\theta_2 - \theta_1)}\right)$$

*Equation 49*

Equation 48, Equation 49 may be combined into a single equation:

$$\tau_0 \frac{d\theta_0}{dt} + \tan^{-1}\left(\frac{\rho \sin(\theta_0)}{1 + \rho \cos(\theta_0)}\right) = (\omega_2 - \omega_1)\tau_0$$

*Equation 50*

in the phase difference:

$$\theta_0 = \theta_2 - \theta_1$$

*Equation 51*

In principle, given a solution of Equation 50, one can solve Equation 48, Equation 49 to find $\theta_1, \theta_2$.

In the case of weak injection $\rho \ll 1$ Equation 50 reduces to the Adler equation:

$$\tau_0 \frac{d\theta_0}{dt} + \rho \sin(\theta_0) = (\omega_2 - \omega_1)\tau_0$$

*Equation 52*

Consequently, the dual oscillator configurations considered herein may be expected to display similar injection locking and pulling behaviour as a single oscillator under injection by an external source.

### 2.5.2. Phase noise analysis

Attention is now turned to the oscillator phase noise spectral density. The phase fluctuations within the oscillator responsible for oscillator phase noise extend to frequencies at which the slowly varying assumption in the preceding analysis is invalid and the effect of the BPF and sidemode resonances cannot be ignored. However, the magnitude of the fluctuations is small in the frequency range of interest permitting linearization of the evolution equations Equation 36, Equation 37 and the application of a Fourier transform.

The linearization of the injection phase shift:

$$\tan^{-1}\left(\frac{\rho \sin(\theta)}{1 + \rho \cos(\theta)}\right) \to \eta\theta \quad ; \quad \eta = \frac{\rho}{1 + \rho}$$

*Equation 53*

is found to be accurate in a large neighborhood of $\theta = 0$. Hence for zero detuning the linearization of Equation 36 and Equation 37 is given by the $2 \times 2$ system:

$$\Theta = \mathcal{D}(\Phi + \mathcal{C}\Theta)$$

*Equation 54*

where:

# $\mathcal{PT}-$symmetric cross-injection dual optoelectronic oscillator

$$\boldsymbol{\Theta} = \begin{bmatrix} \theta_1 \\ \theta_2 \end{bmatrix}$$

*Equation 55*

is the oscillator phase-model state vector:

$$\boldsymbol{\Phi} = \begin{bmatrix} \phi_1 \\ \phi_2 \end{bmatrix}$$

*Equation 56*

is the tuning phase vector:

$$\mathcal{D} = \begin{bmatrix} h \otimes D_{\tau_1} & 0 \\ 0 & h \otimes D_{\tau_2} \end{bmatrix}$$

*Equation 57*

is an operator describing the action of the combined delay and BPF within each loop; and:

$$\mathcal{C} = \begin{bmatrix} 1-\eta & \eta \\ \eta & 1-\eta \end{bmatrix}$$

*Equation 58*

describes the coupling of the oscillators by cross-injection.

Taking the Fourier transform:

$$\widehat{\boldsymbol{\Theta}} = \widehat{\mathcal{D}}(\widehat{\boldsymbol{\Phi}} + \mathcal{C}\widehat{\boldsymbol{\Theta}})$$

*Equation 59*

where:

$$\widehat{\mathcal{D}} = \begin{bmatrix} D_1 & 0 \\ 0 & D_2 \end{bmatrix} \quad ; \quad \begin{cases} D_1(\omega) = H_1(\omega)\exp(-i\omega\tau_1) \\ D_2(\omega) = H_2(\omega)\exp(-i\omega\tau_2) \end{cases}$$

*Equation 60*

yields a formal solution:

$$\widehat{\boldsymbol{\Theta}} = (\mathbf{I} - \widehat{\mathcal{D}}\mathcal{C})^{-1}\widehat{\mathcal{D}}\widehat{\boldsymbol{\Phi}}$$

*Equation 61*

Performing the matrix inverse yields:

$$\widehat{\boldsymbol{\Theta}} = \frac{1}{D_0}\begin{bmatrix} [1-(1-\eta)D_2]D_1 & \eta D_1 D_2 \\ \eta D_2 D_1 & [1-(1-\eta)D_1]D_2 \end{bmatrix}$$

*Equation 62*

where:

$$D_0 = [1-(1-\eta)D_1][1-(1-\eta)D_2] - \eta^2 D_1 D_2$$

*Equation 63*

The linearized model of an injection-locked oscillator is equivalent to a controlled oscillator with proportional gain $\eta$. The cross-injection-locked oscillators act as linear systems arranged in a ring (i.e., the figure of eight loop) that filter the phase fluctuations at their input (the injection port). This motivates rearranging Equation 62, Equation 63 into the form:

$$\widehat{\boldsymbol{\Theta}} = \frac{1}{1-\eta^2 \mathcal{H}_1 \mathcal{H}_2}\begin{bmatrix} \mathcal{H}_1 & \eta \mathcal{H}_1 \mathcal{H}_2 \\ \eta \mathcal{H}_2 \mathcal{H}_1 & \mathcal{H}_2 \end{bmatrix}\widehat{\boldsymbol{\Phi}}$$

*Equation 64*

$$\mathcal{H}_1 = \frac{D_1}{1-(1-\eta)D_1}$$

*Equation 65*

# $\mathcal{PT}$−symmetric cross-injection dual optoelectronic oscillator

$$\mathcal{H}_2 = \frac{D_2}{1 - (1-\eta)D_2}$$

*Equation 66*

The free oscillation of an oscillator under injection is normally quenched. However, when locked the transfer functions $\mathcal{H}_1, \mathcal{H}_2$ share a common resonance for which the loop gain $\eta^2 \mathcal{H}_1 \mathcal{H}_2$ is unity[2]. The scalar function that pre-multiplies the matrix on the right-hand side of Equation 64 consequently contains the pole at the frequency origin characteristic of a free oscillator. The component oscillators of the dual oscillator are locked to each other, but the dual oscillator is free.

In a small neighbourhood of a resonance well within the passbands of the BPF the transfer functions $\mathcal{H}_1$, $\mathcal{H}_2$ may be approximated by:

$$\mathcal{H}_1(\omega) \cong \frac{1}{\eta} \frac{1}{1 + i(\omega - \omega_{p_1})(\tau_1/\eta)} \quad ; \quad \omega_{p_1} \tau_1 = 2p_1 \pi \quad p_1 \in \mathbb{Z}$$

*Equation 67*

$$\mathcal{H}_2(\omega) \cong \frac{1}{\eta} \frac{1}{1 + i(\omega - \omega_{p_2})(\tau_2/\eta)} \quad ; \quad \omega_{p_2} \tau_2 = 2p_2 \pi \quad p_2 \in \mathbb{Z}$$

*Equation 68*

Weak injection results in a narrow bandwidth large magnitude resonance and, conversely, strong injection suppresses and broadens the resonance.

To gain intuition Equation 64 is re-expressed in a form:

$$\widehat{\Theta} = \frac{1}{1 - \eta^2 \mathcal{H}_1 \mathcal{H}_2} \begin{bmatrix} \mathcal{H}_1 & 0 \\ 0 & \mathcal{H}_2 \end{bmatrix} \begin{bmatrix} 1 & \eta\mathcal{H}_2 \\ \eta\mathcal{H}_1 & 1 \end{bmatrix} \widehat{\Phi}$$

*Equation 69*

suited to the interpretation of the dual oscillator architecture as an outer ring within which the two component oscillators are embedded as two inner rings with phase transfer functions $\eta\mathcal{H}_1, \eta\mathcal{H}_2$. Figure 5 provides an example of a numerical evaluation of the phase noise spectral density derived from Equation 69 along with intermediate results. The vector $\widehat{\Phi}$ represents the intrinsic phase fluctuation within the two inner rings. In the example, the component phase fluctuations are taken to be independent but identically distributed stochastic processes characterised by a power law spectral density:

$$S_{\phi\phi}(f) = c_{-2} f^{-2} + c_{-1} f^{-1} + c_0 \quad rad^2/Hz$$

$$c_{-2} = 1.2 \times 10^{-9} \; rad^2 Hz$$
$$c_{-1} = 1.2 \times 10^{-11} \; rad^2$$
$$c_0 = 4 \times 10^{-16} \; rad^2/Hz$$

*Equation 70*

where $f$ is the offset frequency ($Hz$). The phase fluctuation spectral density defined by Equation 70 is plotted in Figure 5(a). The coefficients of the power law are chosen to be representative of experimental observations and to reproduce the spectral densities generated by the phase fluctuation blocks used in the simulations (see section3).

---

[2] This is ensured by the gain saturation mechanism. For simplicity in the phase noise analysis, it is assumed that the passband centre frequency of the bandpass filter coincides with the common resonance responsible for oscillation.

# $\mathcal{PT}-$symmetric cross-injection dual optoelectronic oscillator

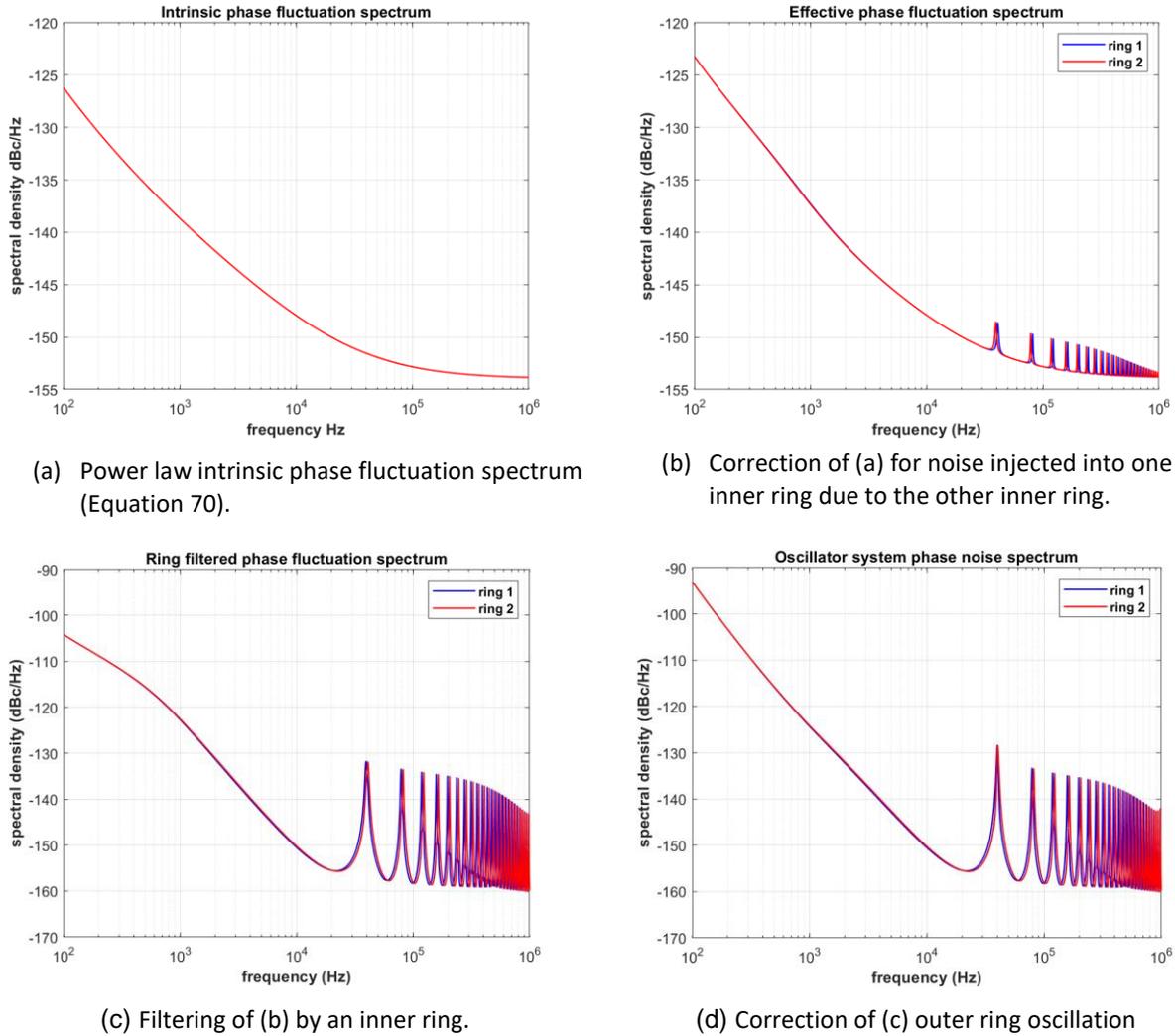

(a) Power law intrinsic phase fluctuation spectrum (Equation 70).

(b) Correction of (a) for noise injected into one inner ring due to the other inner ring.

(c) Filtering of (b) by an inner ring.

(d) Correction of (c) outer ring oscillation

*Figure 5. Construction of the phase noise spectrum of the cross-injection dual TDO predicted by Equation 96 with Equation 70. There is a Vernier effect detuning of the ring resonances due to their distinct free spectral ranges that increases with frequency in the first half of the frequency range and then decreases to zero in the second half of the frequency range. The sidemode revival at 1 MHz that would otherwise occur is suppressed by the RF resonators.*

The product of the rightmost matrix of Equation 69 and $\widehat{\Phi}$ then represents the effective phase fluctuations spectral density (Figure 5 (b)), i.e., the sum of the intrinsic phase fluctuations within each ring $\hat{\phi}_1$, $\hat{\phi}_2$ and the phase fluctuations injected into each ring by the other inner ring $\eta\mathcal{H}_2\hat{\phi}_2$, $\eta\mathcal{H}_1\hat{\phi}_1$. The resonances of $\eta\mathcal{H}_1$ and $\eta\mathcal{H}_2$ have peaks of magnitude of unity and troughs of magnitude of the order of $\eta$. The bandwidth of the resonances is proportional to the injection parameter $\eta$. Consequently, for weak injection the contribution by the intrinsic phase fluctuations is the predominant component of the effective phase fluctuation spectral density. The injected component creates, for uncorrelated intrinsic fluctuations, small narrow bumps of at most 3 dB in magnitude in the spectral density. Moreover, these bumps are detuned from almost all the resonances of the inner ring.

The central diagonal matrix describes the action of the rings as phase filters. Under injection the resonances falling within the passband of the BPF have peaks limited in magnitude to $1/\eta$. The sidemodes are substantially suppressed by injection compared to the free oscillators ($\eta = 0$) where sidemode resonances are damped only by the BPF (Figure 5 (c)). Nevertheless, it is this term in the

# $\mathcal{PT}-$symmetric cross-injection dual optoelectronic oscillator

decomposition that is principally responsible for the sidemode resonance structure observed in the phase noise spectral density of the outer ring. Only the singlet sidemode resonances associated with the respective delay are seen in the phase noise spectrum of the output of each inner ring. However, this is a consequence only of the choice of measurement. The outputs of the component oscillators of the $SU(1,1)-$ configuration may be combined by a coupler with a Hadamard transmission matrix to observe the doublet structure of the $\mathcal{PT}-$symmetric configuration (Figure 11 (b)).

The leftmost scalar function accounts for the oscillation of the outer ring at a common resonance of its two component rings ensured by the locking dynamics by re-introducing a pole at the frequency origin characteristic of a free oscillator. The outer ring oscillation drives the component oscillators quenching their free oscillation and suppressing their sidemodes. In a small neighbourhood of the frequency origin, Equation 69 may be reduced to:

$$\hat{\theta}_1(\omega) = \hat{\theta}_2(\omega) \cong \frac{1}{i\omega\tau}\hat{\phi}(\omega) \quad \begin{cases} \tau = \tau_1 + \tau_2 \\ \hat{\phi} = \hat{\phi}_1 + \hat{\phi}_2 \end{cases}$$

*Equation 71*

The phase fluctuation spectral densities $\hat{\theta}_1, \hat{\theta}_2$ are identical at low offset frequencies, i.e., the two oscillators are rigidly locked in this spectral region. Treated as a single oscillator, the dual oscillator has a delay equal to the sum of the delays of its components but is subject to intrinsic phase fluctuations also equal to the sum of the intrinsic phase fluctuations of its components. If the oscillators have comparable delays and comparable but uncorrelated low-frequency internal fluctuations, then the close-in phase noise is reduced by 3 dB compared to a single oscillator with similar delay.

The Vernier effect ensures there will be periodic revivals of the sidemodes at frequencies in a neighbourhood $\omega_p\tau_d \equiv 0 \bmod 2\pi$ where $\eta^2\mathcal{H}_1(\omega)\mathcal{H}_2(\omega)$ can approach unity if the revival occurs within the passband of the BPF. For sufficiently small $\tau_d$ the revivals may be substantially suppressed by the transfer function $H$ of the BPFs.

| | |
|---|---|
| **Single loop reference oscillator** | |
| $\tau_D = 24.9\ \mu s$ | *Delay line delay time* |
| $\tau_R = 0.1\ \mu s$ | *Resonator on-resonance group delay* |
| $\Delta f = 1/\pi\tau_R = 3.18\ MHz$ | *Resonator bandwidth* |
| $\tau_G = 25\ \mu s$ | *Round trip group delay time* |
| $FSR = 1/\tau_G = 40\ kHz$ | *Frequency interval between modes* |
| **$\mathcal{PT}-$ symmetric model 1 oscillator** | |
| $\tau_1 = \tau_2 = 24.4\ \mu s$ | *Dual matched delay line delay time* |
| $\tau_R = 0.1\ \mu s$ | *Resonator on-resonance group delay* |
| $\tau_\alpha = \tau_\beta = 0.5\mu s$ | *MZI A & B delay times* |
| $\tau_\alpha + \tau_\beta = 0.5\ \mu s$ | *Common part of coupler delay* |
| $\gamma = 0.126$ | *Gain/loss parameter* |
| **$\mathcal{PT}-$ symmetric model 2 oscillator** | |
| $\tau_1 = 25.4\ \mu s$ | *Loop 1 delay time* |
| $\tau_2 = 24.4\ \mu s$ | *Loop 2 delay time* |
| $\tau_R = 0.1\ \mu s$ | *Resonator on-resonance group delay* |
| $\rho = \tanh(\gamma) = 0.125$ | *Cross injection ratio* |
| $\eta = \rho/(1 + \rho) = 0.1$ | *Cross injection parameter* |

*Table 1 Principal simulation model parameters*

## 3. Simulation

### 3.1. Simulink™ models

To verify the theoretical analysis, Simulink models have been developed that simulate the $\mathcal{PT}-$symmetric TDO in the time domain. Frequency domain results such as RF and phase noise spectral densities are obtained by spectral analysis of the time domain signals. The models are composed of blocks that model oscillator components parts that are arranged according to the circuit architecture. The Simulink models follow the development of the circuit architecture presented in Section 2. The principal model parameters are listed in Table 1.

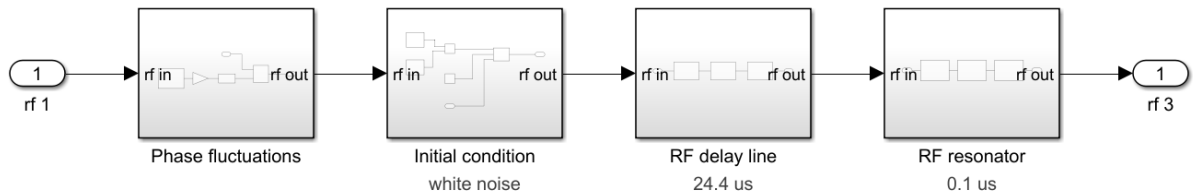

(a) Passive loop 1 & 2 subsystem. The total loop delay time is 24.5 µs to which the common part of the coupler delay adds $0.5$ µs to give the round-trip time 25 µs

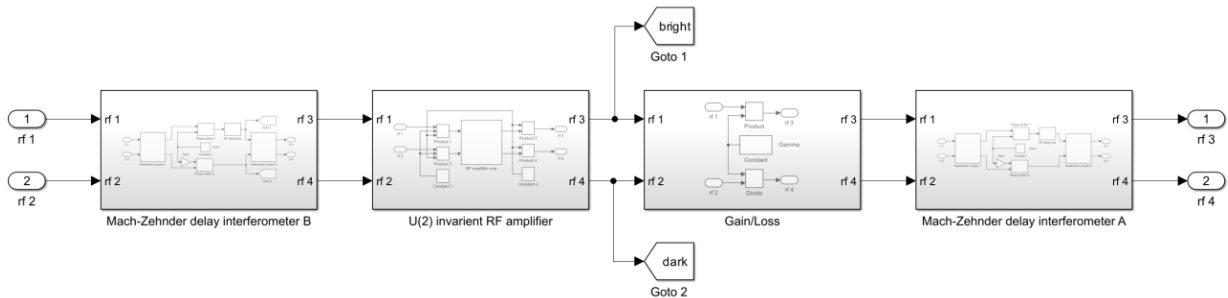

(b) Active $\mathcal{PT}-$symmetric coupler subsystem.

*Figure 6. $\mathcal{PT}-$symmetric model 1 oscillator detail of the two loops and coupler.*

An envelope model represents an analytic signal by the product of a complex envelope and a pure carrier with specified nominal frequency. Since the carrier is completely known it conveys no information. Consequently, it is only necessary to model the induced map between the incoming and outgoing complex envelope to fully describe a component. The envelope representation is motivated by the vastly reduced sample rate required to avoid aliasing in discrete time simulations of narrow-band signals with large carrier frequencies.

Envelope models may be reduced to phase-only models that are useful for modelling the evolution of an established oscillation as in the injection locking analysis presented in Section 2.5. However, implicit in the assumption of zero amplitude fluctuations is operation in deep saturation. Consequently, the linear regime of the initial transient is not modelled, limiting the initial condition to an established oscillation. It is also common to invoke the Leeson phase-only model of an RF resonator for reasons of its simplicity and utility. However, it has been found that the Leeson model can fail in specific injection pulling regimes. For these reasons, reduced phase-only models are not used in the simulations presented in this work.

# $\mathcal{PT}-$symmetric cross-injection dual optoelectronic oscillator

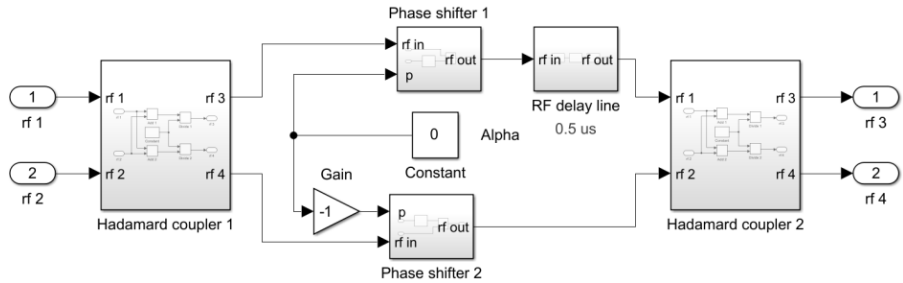

Figure 7. Detail of Mach-Zehnder delay interferometer A (Figure 6 (b)). Interferometer B has identical structure with parameter $\alpha$ substituted by $\beta$. The linear in frequency differential phase shift required for a mode-selective $\mathcal{PT}-$symmetric phase transition is provided by the RF delay line block.

Simulation model 1 is a literal implementation of a $\mathcal{PT}-$symmetric dual TDO with a frequency dependent symmetry-breaking phase transition, i.e., it is mode selective. It consists of two passive loops with identical delays that connect the output ports of an active coupler subsystem to its input ports to form two coupled ring oscillators as in Figure 1 (a). The passive loops are each composed of four Simulink subsystem blocks that model the RF resonator and the RF delay line, its white noise initial condition and power-law phase fluctuations, as shown in Figure 6. The block parameters are identical in each loop, but the noise sources are independent. The active coupler contains a dual matched saturating amplifier and a pair of linear amplifiers providing matched gain and loss. The amplifiers are placed between a pair of MZIs that act as variable couplers controlled by the differential phase shift of the interferometer arms (see Figure 7). The differential phase shift is composed of a static bias set by the $\alpha_0$ or $\beta_0$ parameters and a component linear in frequency set by a short delay line with delay time $2\tau_\alpha$ or $2\tau_\beta$ in one arm of the interferometer.

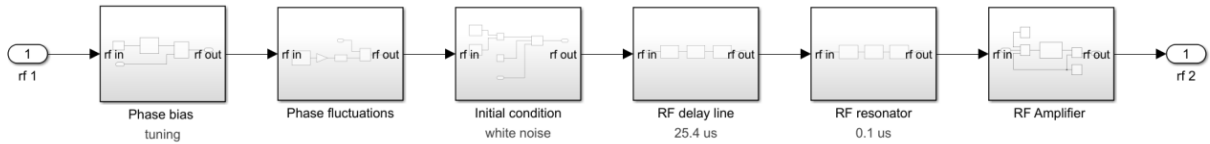

(a) Active loop 1 subsystem. The total delay time is 25.5 μs. The passive loop 2 subsystem has the same structure, but the total delay time is 24.5 μs.

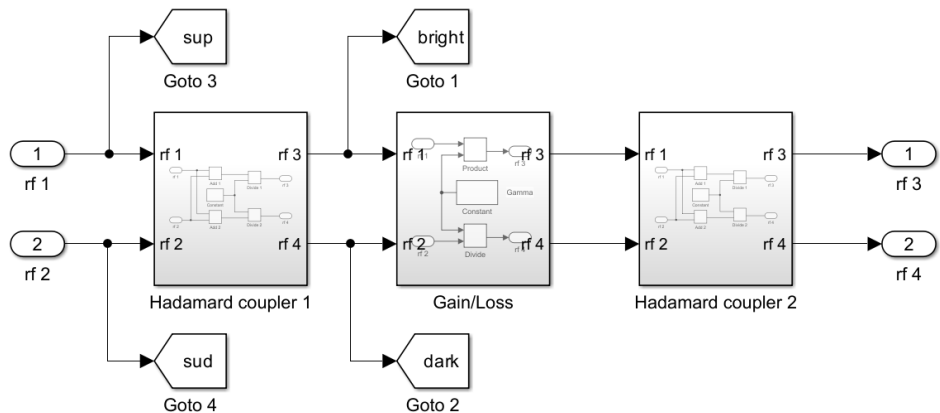

(b) Real SU(1,1) coupler subsystem

Figure 8. $\mathcal{PT}-$symmetric model 2 oscillator detail of the two loops and coupler.

# $\mathcal{PT}-$symmetric cross-injection dual optoelectronic oscillator

Simulation model 2 is derived by annihilation of the outermost Hadamard couplers of the Mach-Zehnder delay interferometer A & B of model 1 (see Figure 3) via their commutation with the pair of identical loops (see Figure 7). The interferometer arms then concatenate with the identical loops of model 1 to form the two unbalanced loops of model 2. The $U(2)-$invariant RF amplifier and the Hadamard coupler that precedes it commute and are interchanged, thereby forming the active real $SU(1,1)$ coupler from the gain/loss block with the innermost Hadamard coupler remnants of the interferometers placed on either side. Simulations confirm the equivalence of this configuration and model 1. The new location of the $U(2)$ invariant RF amplifier enables its replacement by matched independent limiting RF amplifiers without change to the outcome of the initial mode competition or the phase noise spectral density of the established oscillation. Finally, model 2 results from incorporating the limiting RF amplifiers into the two loops coupled by the remaining $SU(1,1)$ coupler block with contents shown in Figure 8(b). Model 2 is found to be robust to parameter variations. Strict matching of the RF amplifiers is not necessary even though this may break the desired linear regime symmetry, unbalance the cross-injection ratios, and somewhat prolong evolution to a locked state. The resultant dual TDO with symmetric cross-injection is consequently the most suited to hardware implementation.

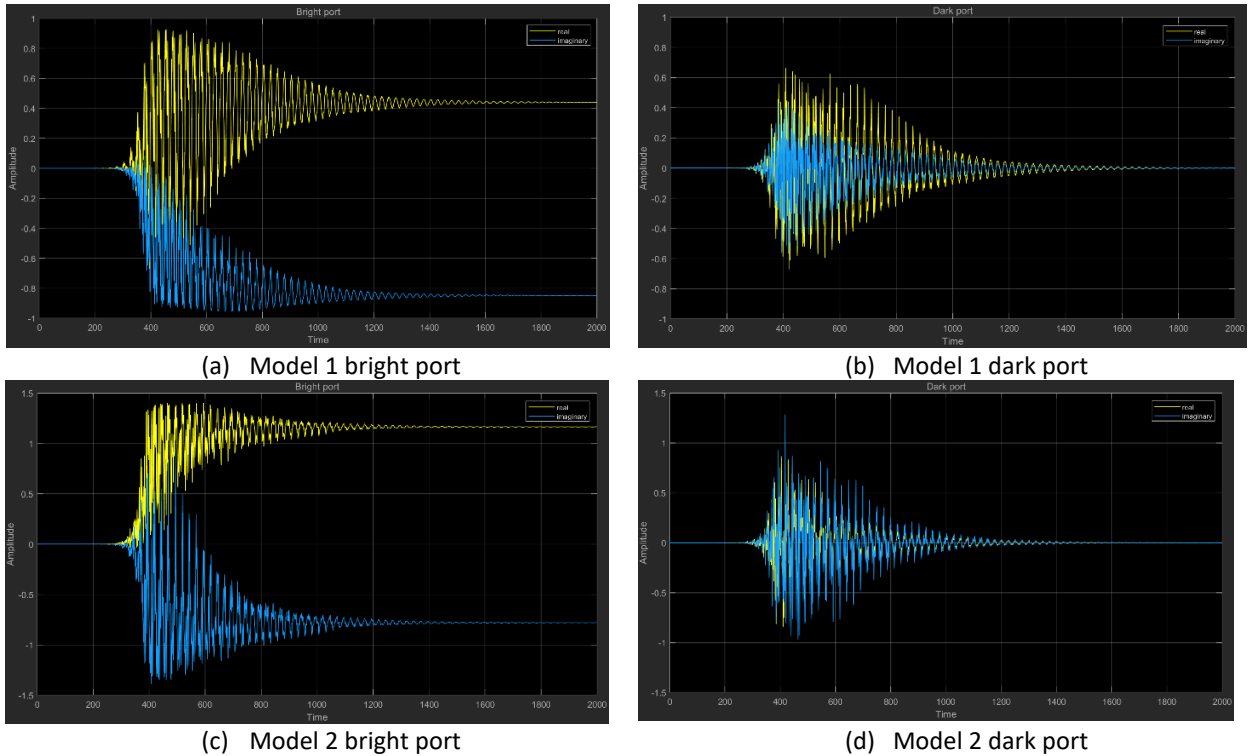

(a) Model 1 bright port  (b) Model 1 dark port
(c) Model 2 bright port  (d) Model 2 dark port

*Figure 9. $\mathcal{PT}-$ symmetric oscillator simulation model 1 & 2 bright & dark port complex envelope initial transient. Time in 1 µs units.*

Extensive simulations have been performed to validate the theory of the $\mathcal{PT}-$ symmetric dual TDO including the equivalence of the two configurations (simulation models 1 & 2) and the predicted phase noise spectral density. Figure 9, Figure 10, Figure 11, Figure 12, and Figure 13 present a representative sample of results.

# $\mathcal{PT}-$symmetric cross-injection dual optoelectronic oscillator

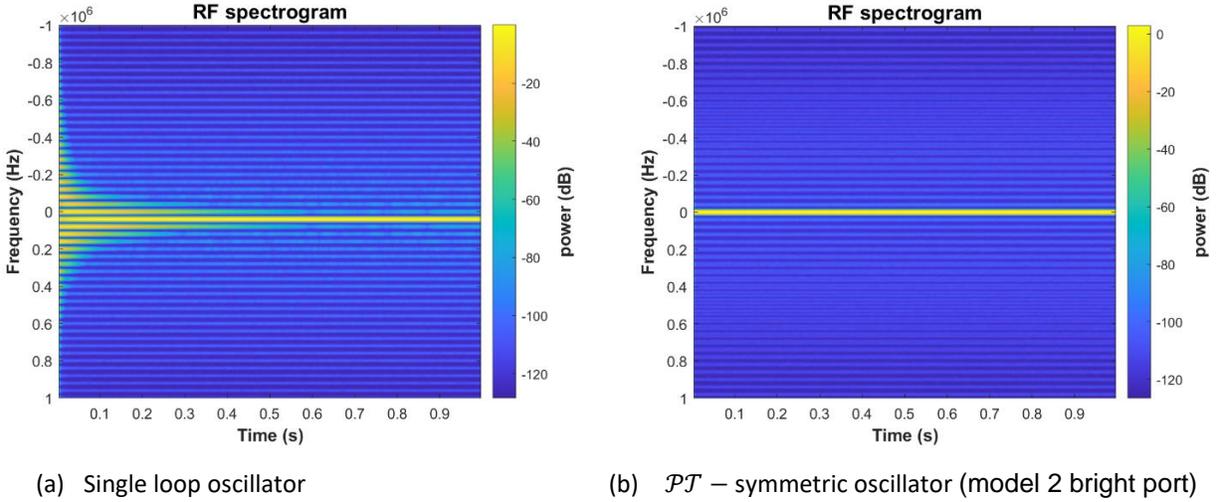

(a) Single loop oscillator
(b) $\mathcal{PT}-$ symmetric oscillator (model 2 bright port)

*Figure 10. Comparison of the evolution towards a single mode oscillating state of the $\mathcal{PT}-$ symmetric oscillator (model 2) and a single loop oscillator with the same phase fluctuations & initial conditions parameters and similar round-trip delay (25 μs). The single loop oscillator requires ~500 ms for its sidemodes to decay to a noise-driven residual level, while the ~2ms initial transient of the $\mathcal{PT}-$ symmetric oscillator is too brief to be visible in the RF spectrogram.*

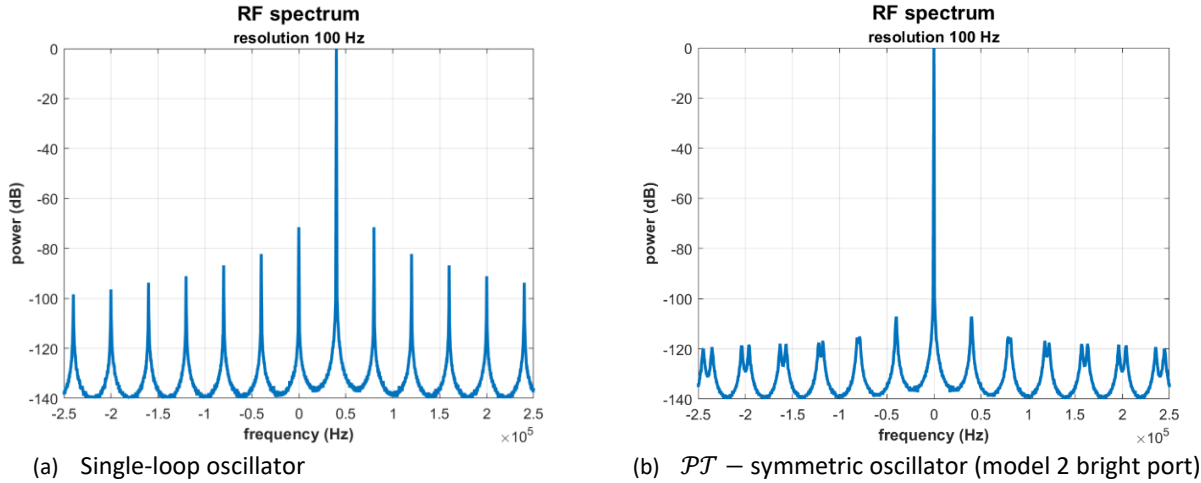

(a) Single-loop oscillator
(b) $\mathcal{PT}-$ symmetric oscillator (model 2 bright port)

*Figure 11. Comparison of the asymptotic RF spectral density of the $\mathcal{PT}-$ symmetric oscillator (model 2) and a single loop oscillator with the same phase fluctuations & initial conditions parameters and similar round-trip delay (25 μs). The unbroken $\mathcal{PT}-$ symmetry doublet resonances are clearly visible in the RF spectral density of the $\mathcal{PT}-$ symmetric oscillator. In the case of the single-loop oscillator the mode competition process starting from noise has resulted in an oscillating mode displaced from the RF resonator passband centre frequency by one FSR while the mode-selective $\mathcal{PT}-$ symmetry phase transition results in an oscillating mode at passband centre.*

Figure 9 provides time-domain plots of the start-up phase from low-level noise to established oscillation. The mode-selective $\mathcal{PT}-$symmetry phase transition is remarkably effective at accelerating the establishment of single mode oscillation within ~2 ms. The oscillation initially builds up in both the gain path (bright port) and loss paths (dark port) but the loss path oscillation soon decays leaving a gain path oscillation with a substantially constant complex envelope characteristic of oscillation of the main mode at zero-offset frequency. Figure 10 presents two spectrograms (short time Fourier transforms) that contrast the evolution of the RF spectrum of the oscillations of a single loop oscillator and the $\mathcal{PT}-$ symmetric oscillator. The sidemodes of the initial oscillation of the single-loop oscillator decay slowly; a

# $\mathcal{PT}-$symmetric cross-injection dual optoelectronic oscillator

good proportion of a second elapses before the sidemodes reach a residual level driven by noise. The regular frequency interval between modes is clearly visible. In contrast, the initial transient of the $\mathcal{PT}-$symmetric oscillator is too brief to be resolved by the spectrogram which appears visually uncorrupted by noise.

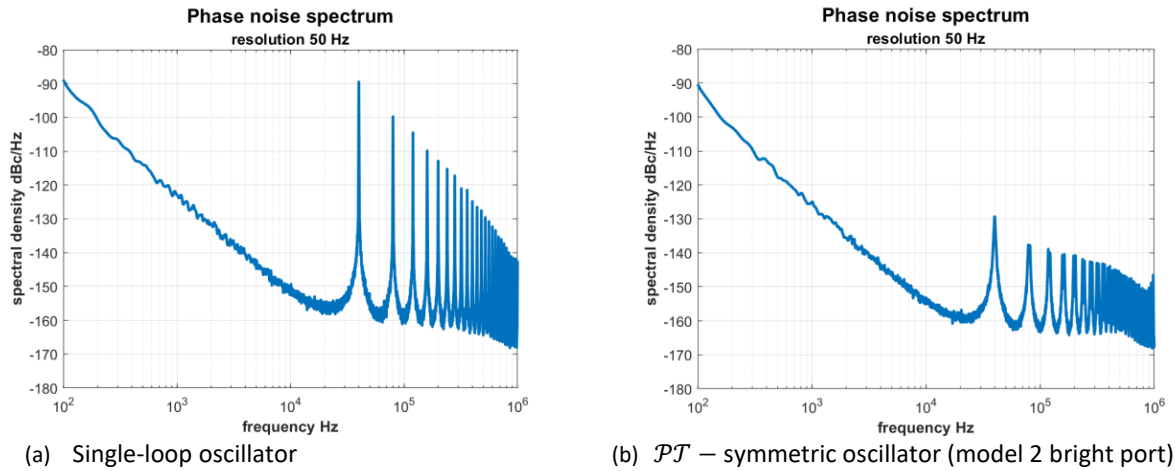

(a) Single-loop oscillator  (b) $\mathcal{PT}-$ symmetric oscillator (model 2 bright port)

*Figure 12. Comparison of the phase noise spectral density of the $\mathcal{PT}-$ symmetric oscillator (model 2) and a single loop oscillator with the same phase fluctuation & initial condition parameters and similar round-trip delay (25 µs). The $\mathcal{PT}-$ symmetric oscillator spurious sidemodes are significantly suppressed relative to the single-loop oscillator.*

The characteristic doublet structure of the PT-symmetric phase and the slow periodic variation of the doublet splitting frequency are just visible in Figure 10. These characteristic distinguishing features of the single mode and oscillator and the $\mathcal{PT}-$ symmetric oscillator are clearer in the RF spectrum of the respective oscillation shown in Figure 11. It can also be observed in the case of the single-loop oscillator that the mode-competition process has selected a mode displaced by one FSR from the intended main mode at zero offset frequency, while the $\mathcal{PT}-$ symmetric oscillator systematically selects the correct mode that is far more favoured by its broken $\mathcal{PT}-$ symmetry.

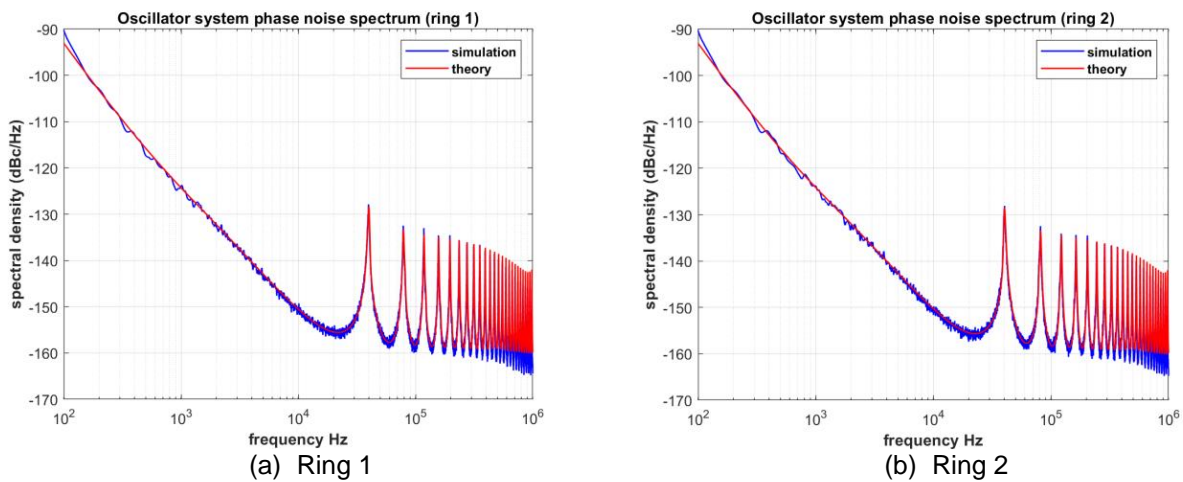

(a) Ring 1  (b) Ring 2

*Figure 13. Comparison of the phase noise spectral density generated using data generated by simulation model 2 with data predicted analytically (Equation 96 with Equation 70.). The departure at frequencies approaching 1 MHz of the simulation data from the theoretical fit is an aliasing artefact clearly visible in Figure 12 (b). The sample interval is $0.5$ µs.*

# $\mathcal{PT}-$symmetric cross-injection dual optoelectronic oscillator

Figure 12 presents a comparison of the phase noise spectral density of the data generated by the single loop and $\mathcal{PT}-$ symmetric oscillator simulation models. It is observed that both oscillators have comparable phase noise at offset frequencies below the first sidemode resonance but the sidemode level of the single loop oscillator is substantially greater (~ 60 dB for the first sidemode) than the sidemode level of the $\mathcal{PT}-$ symmetric oscillator. Figure 13 presents a comparison of the spectral density of the data generated by ring 1 and ring 2 ports of the $\mathcal{PT}-$ symmetric oscillator and the analytic prediction of the phase only injection locking theory (Equation 96 with Equation 70.). The agreement between theory and simulation is excellent.

## 4. Summary & Conclusions

The circuit architectures and accompanying theory presented in the prior literature on $\mathcal{PT}-$ symmetric OEOs have been found to have deficiencies that cast doubt on the claimed observations of a $\mathcal{PT}-$ symmetry phase transition. A serious flaw is the projection from a 2 dimensional to 1 dimensional oscillator state variable in some segment of the oscillating loop, no doubt in the interests of hardware economy. It has been shown that the reduction of dimension eliminates the $\mathcal{PT}-$ symmetry phase transition effectively reducing the architecture to a standard single loop OEO in the case of optical delay lines of the same length or a standard dual loop OEO in the case of optical delay lines of dissimilar length. Moreover, it has been shown that without some mechanism to vary the gain/loss or coupling with frequency, the $\mathcal{PT}-$ symmetry phase transition is not mode selective. Finally, the theory invoked to explain the expected phenomena is only valid in the brief linear regime and assumes adiabatic energy exchange between two resonators, i.e., weak coupling. While the experimental results are captured in the deep saturation regime where the oscillation is fully established. Consequently, any connection between the theory, circuit architecture and measurements in the prior literature is found to be tenuous.

This paper corrects all these deficiencies. A delay differential / integral equation is used to model a $\mathcal{PT}-$ symmetric oscillator. Scattering matrix methods are used to construct the $\mathcal{PT}-$ invariant transmission matrix characterising the coupler. The coupler transmission matrix is found to be necessarily invertible, thereby preserving the dimension of the state variable, and no assumption of weak coupling is made. The $\mathcal{PT}-$ symmetric oscillator circuit architecture thereby identified can be implemented using standard RF-photonic components except for the sustaining amplifier which requires a particular saturation mechanism to preserve $\mathcal{PT}-$ invariance. A mode-selective $\mathcal{PT}-$ symmetry phase transition is achieved by placing a short delay within one arm of a Mach-Zehnder interferometer that appears within the circuit architecture.

In addition, the mode-selective $\mathcal{PT}-$ symmetric oscillator has been found to possess an alternative but equivalent configuration as a cross-injection dual TDO with similar but distinct delays. The local $\mathcal{PT}-$ symmetry phase transition may then be understood in terms of the Vernier effect. The cross-injection perspective facilitates the extension of the theory to a phase-only model of an established oscillation that inter alia provides an analytic prediction of the phase-noise spectral density. Advantageously, the alternative configuration enables the special sustaining amplifier to be replaced by pair of matched but otherwise independent amplifiers. Thereby, the second configuration is amenable to practical implementation as a dual OEO using standard RF-photonic and RF-electronic components throughout. Both configurations of the circuit architecture map neatly and expressively into a Simulink™ simulation model. Extensive simulation trials have been performed to validate the theory, observe known $\mathcal{PT}-$ symmetry breaking phenomena, confirm the equivalence of the two configurations, and reproduce the phase-noise spectral density predicted by the analytic theory. There is excellent agreement between the theoretical and simulation results.

This work is valuable not only in advancing the theory of $\mathcal{PT}-$ symmetric TDOs in general and but also in demonstrating the potential of a $\mathcal{PT}-$ symmetric OEO to offer applications the desirable attribute of combined low phase noise and high sidemode suppression.

## 6. Appendix I Scattering matrices & $\mathcal{PT}$ − symmetry

A LTI system containing one or more components or devices is fully characterised by a scattering matrix. The scattering matrix is a complex matrix-valued function of frequency $\omega$ that describes the scattering of electromagnetic energy from an indexed set of incoming modes to an indexed set of outgoing modes. To every incoming propagating mode there is an associated outgoing propagating mode which is a time-reversed replica of the inward propagating mode and vice versa. The scattering matrix $\mathbf{S}(\omega) \in \mathbb{C}^{n \times n}$ maps a vector of complex amplitudes $\boldsymbol{a}(\omega) \in \mathbb{C}^n$ of the set of incoming modes to a vector of complex amplitudes $\boldsymbol{b}(\omega) \in \mathbb{C}^n$ of the set of outgoing modes.

$$\boldsymbol{b} = \mathbf{S}\boldsymbol{a}$$

*Equation 72*

$$\mathbf{S} = \begin{bmatrix} S_{11} & \cdots & S_{1n} \\ \vdots & \ddots & \vdots \\ S_{n1} & \cdots & S_{nn} \end{bmatrix} \quad ; \quad \boldsymbol{a} = \begin{bmatrix} a_1 \\ \vdots \\ a_n \end{bmatrix} \quad ; \quad \boldsymbol{b} = \begin{bmatrix} b_1 \\ \vdots \\ b_n \end{bmatrix}$$

*Equation 73*

It is emphasized that the scattering matrix is a Fourier domain object. The system is described in the time domain by a matrix of convolutional operators. However, if the scattering matrix varies little over the spectral extent of the incoming modes, it may be approximated by the constant matrix $\mathbf{S}(\omega_0)$ where $\omega_0$ is the nominal centre frequency of the incoming modes. The matrix $\mathbf{S}(\omega_0)$ may then be taken out of the Fourier integral and applied in the time domain as an approximation to a wideband system. The nominal frequency $\omega_0$ is often suppressed for simplicity of notation.

### Energy conservation

If the system conserves energy, that is there are no other sources (pumps) or sinks (dissipation) other than the modes included in the scattering matrix, then:

$$\|\boldsymbol{b}\|^2 = (\boldsymbol{b}, \boldsymbol{b}) = (\mathbf{S}\boldsymbol{a}, \mathbf{S}\boldsymbol{a}) = (\mathbf{S}^\dagger \mathbf{S}\boldsymbol{a}, \boldsymbol{a}) = (\boldsymbol{a}, \boldsymbol{a}) = \|\boldsymbol{a}\|^2$$

*Equation 74*

where $(\cdot,\cdot)$ and $\|\cdot\|$ are the standard inner product and its induced norm on $\mathbb{C}^n$.

It follows that the scattering matrix is then *unitary*:

$$\mathbf{S}^\dagger \mathbf{S} = \mathbf{I}$$

*Equation 75*

### Reciprocity

Electromagnetic scattering problems generally obey Lorentz reciprocity, i.e., if incoming mode $j$ has amplitude $a_j$ and is scattered into outgoing mode $k$ having amplitude $b_k$, then, if incoming mode $k$ has amplitude $a_k = b_k$ it is scattered into the outgoing mode $j$ with amplitude $b_j = a_j$. That is $s_{jk} = s_{kj}$ and the scattering matrix is *symmetric*.

$$\mathbf{S} = \mathbf{S}^\mathrm{T}$$

*Equation 76*

Suppose the ports of the system are partitioned into two disjoint sets which may be considered as a set of ingress ports and a set of egress ports. Then the scattering matrix partitions in the same way:

$$\begin{bmatrix} \boldsymbol{b}_1 \\ \boldsymbol{b}_2 \end{bmatrix} = \begin{bmatrix} \mathbf{S}_{11} & \mathbf{S}_{12} \\ \mathbf{S}_{21} & \mathbf{S}_{22} \end{bmatrix} \begin{bmatrix} \boldsymbol{a}_1 \\ \boldsymbol{a}_2 \end{bmatrix}$$

*Equation 77*

That is:

$S_{11}$ describes the backscattering (reflection) of incoming modes at the ingress ports into outgoing modes at the ingress ports.

$S_{12}$ describes the forward-scattering (transmission) of incoming modes at the egress ports to outgoing modes at the ingress ports.

$S_{21}$ describes the forward-scattering (transmission) of incoming waves at the ingress ports to outgoing modes at the egress ports.

$S_{22}$ describes the backscattering (reflection) of incoming modes at the egress ports to outgoing modes at the egress ports.

Reciprocity implies a symmetric scattering matrix and consequently:

$$\mathbf{S}_{11} = \mathbf{S}_{11}^{\mathrm{T}} \quad ; \quad \mathbf{S}_{12} = \mathbf{S}_{21}^{\mathrm{T}} \quad ; \quad \mathbf{S}_{22} = \mathbf{S}_{22}^{\mathrm{T}}$$

*Equation 78*

Generally, optical amplifiers are reciprocal while electronic amplifiers are not. Consequently, unidirectional oscillation is natural for electronic oscillators, but bidirectional oscillation is natural for optical oscillators (lasers); it is necessary to insert non-reciprocal isolators within an optical ring oscillator to assert unidirectional operation. For analysis, it is convenient to treat active components that provide either gain or loss as reciprocal. From this perspective an electronic amplifier might be composed of two matched non-reciprocal amplifiers, one operating in the forward direction and the other operating in the reverse direction, aided by directional couplers and isolators or circulators. This conceptual device is of no practical consequence as conventional amplifiers may substitute for the reciprocal amplifiers as appropriate to the desired direction of operation.

## Transmission submatrices

The partition is most useful for the reduced problem in which backscattering is negligible:

$$\mathbf{S}_{11} \sim \mathbf{0} \quad ; \quad \mathbf{S}_{22} \sim \mathbf{0}$$

*Equation 79*

A unitary scattering matrix then implies:

$$\mathbf{S}_{12}^{\dagger} \mathbf{S}_{12} \sim \mathbf{I} \quad ; \quad \mathbf{S}_{21}^{\dagger} \mathbf{S}_{21} \sim \mathbf{I}$$

*Equation 80*

The most important findings are that the forward and reverse transmission matrices $\mathbf{S}_{21}$, $\mathbf{S}_{12}$ are the transpose of each other under reciprocity and inherit the unitary property of the full scattering matrix in the absence of loss (sinks) or gain (sources) provided backscattering is negligible; but they are not necessarily symmetric.

## Reversable reciprocal systems

A common symmetry is that the partitioned system is *reversible*, i.e., the partitioned system is identical under an interchange of ingress and egress ports:

$$\mathbf{S}_{12} = \mathbf{S}_{21} \quad ; \quad \mathbf{S}_{22} = \mathbf{S}_{11}$$

*Equation 81*

That is:

$$\mathbf{S} = \begin{bmatrix} \mathbf{S}_{11} & \mathbf{S}_{12} \\ \mathbf{S}_{12} & \mathbf{S}_{11} \end{bmatrix}$$

*Equation 82*

$\mathcal{PT}-$symmetric cross-injection dual optoelectronic oscillatorIt then follows that the all the submatrices of the partition are *symmetric* under reciprocity.

## Invertible reciprocal systems

Another common symmetry is that the partitioned system is invariant to an *inversion*, i.e., $\mathbf{S}$ is invariant to the conjugation:

$$\mathbf{JSJ} = \mathbf{S}$$

*Equation 83*

by the exchange matri*x* $\mathbf{J}$ which has a trailing diagonal with unit elements and zero elements elsewhere.

Invoking reciprocity (Equation 78) the transfer matrices are *persymmetric*, i.e., symmetric across the trailing diagonal:

$$\mathbf{S}_{12} = \mathbf{JS}_{12}^{\mathrm{T}}\mathbf{J}$$
$$\mathbf{S}_{21} = \mathbf{JS}_{21}^{\mathrm{T}}\mathbf{J}$$

*Equation 84*

The backscattering matrices are symmetric by reciprocity and related by inversion symmetry:

$$\mathbf{S}_{11} = \mathbf{JS}_{22}\mathbf{J}$$

*Equation 85*

but are not persymmetric.

If the partitioned system is *reversable and invertible* then all the submatrices of the partition are *bisymmetric* (i.e., symmetric and persymmetric) under reciprocity.

## Time reversal

Let $u$ be a *real* function of time and $\hat{u}$ its associated Fourier transform:

$$\hat{u}(\omega) = \int_{-\infty}^{\infty} u(t)\exp(-i\omega t)\,dt$$

*Equation 86*

It follows that:

$$\hat{u}^*(\omega) = \int_{-\infty}^{\infty} u(-t)\exp(-i\omega t)\,dt$$

*Equation 87*

Hence, given:

$$\mathbf{b} = \mathbf{Sa}$$

*Equation 88*

under time reversal:

$$\mathbf{a}^* = (\mathbf{S}^{-1})^*\,\mathbf{b}^*$$

*Equation 89*

where $\mathbf{b}^*$ is now the vector of complex amplitudes of *incoming* modes and $\mathbf{a}^*$ is the vector of complex amplitudes of *outgoing* modes. Consequently, the action of *the time reversal* operator $\mathcal{T}$ on $\mathbf{S}$ is given by:

$$\mathbf{S} \xrightarrow{\mathcal{T}} (\mathbf{S}^{-1})^*$$

*Equation 90*

# $\mathcal{PT}-$symmetric cross-injection dual optoelectronic oscillator

Provided backscattering is negligible, the time reversal operator acts on the transmission submatrices and the elements of diagonal matrices similarly.

$$\mathbf{S}_{12} \xrightarrow{\mathcal{T}} (\mathbf{S}_{12}^{-1})^* \quad ; \quad \mathbf{S}_{21} \xrightarrow{\mathcal{T}} (\mathbf{S}_{21}^{-1})^*$$

*Equation 91*

Notably, its action on a unitary matrix reduces to transposition:

$$\mathbf{U} \xrightarrow{\mathcal{T}} (\mathbf{U}^{-1})^* = (\mathbf{U}^\dagger)^* = \mathbf{U}^T$$

*Equation 92*

its action on a diagonal delay matrix with delay time $\tau$ is invariant:

$$\exp(-i\omega\tau) \xrightarrow{\mathcal{T}} \exp(-i\omega\tau)$$

*Equation 93*

and its action on a diagonal gain/loss matrix exchanges gain coefficients with loss coefficients and vice versa:

$$\exp(\gamma) \xrightarrow{\mathcal{T}} \exp(-\gamma)$$

*Equation 94*

## Parity conjugation

The persymmetric transposition:

$$\mathbf{S} \xrightarrow{\mathcal{P}} \mathbf{J}\mathbf{S}^T\mathbf{J}$$

*Equation 95*

is adopted in this work as the definition of the parity conjugation operator $\mathcal{P}$ given its compatibility with the time reversal operation in the special case of a coupled dual TDO with identical delay-line loops with a 4-port coupler with negligible backscattering.

Both $\mathcal{P}$ and $\mathcal{T}$ meet the requirement of an involution by their definition. A further mandatory property is that $\mathcal{P}$ and $\mathcal{T}$ should commute. That is, for any matrix $\mathbf{K} \in \mathbb{C}^{n\times n}$ the operator compositions $\mathcal{P} \circ \mathcal{T}$ and $\mathcal{T} \circ \mathcal{P}$ should agree, which is confirmed by direct calculation:

$$\mathbf{K} \xrightarrow{\mathcal{T}\circ\mathcal{P}} ((\mathbf{J}\mathbf{K}^T\mathbf{J})^{-1})^* = \mathbf{J}(\mathbf{K}^\dagger)^{-1}\mathbf{J}$$

*Equation 96*

$$\mathbf{K} \xrightarrow{\mathcal{P}\circ\mathcal{T}} \mathbf{J}((\mathbf{K}^{-1})^*)^T\mathbf{J} = \mathbf{J}(\mathbf{K}^\dagger)^{-1}\mathbf{J}$$

*Equation 97*

Specialising to $2\times 2$ transmission matrices, any matrix $\mathbf{K} \in \mathbb{C}^{2\times 2}$ admits the singular value decomposition:

$$\mathbf{K} = \mathbf{U}\boldsymbol{\Sigma}\mathbf{V}^T \quad ; \quad \boldsymbol{\Sigma} = \begin{bmatrix} \sigma_1 & 0 \\ 0 & \sigma_2 \end{bmatrix}$$

*Equation 98*

where $\mathbf{U}, \mathbf{V} \in SU(2)$ and $\boldsymbol{\Sigma}$ is a diagonal matrix with the real positive singular values $\sigma_1, \sigma_2$ as its elements. Setting:

$$\kappa = \sqrt{\sigma_1\sigma_2} \quad ; \quad \gamma = \ln\left(\sqrt{\sigma_1/\sigma_2}\right)$$

*Equation 99*

and introducing the unit determinant gain/loss matrix:

# $\mathcal{PT}-$symmetric cross-injection dual optoelectronic oscillator

$$\mathbf{\Gamma}(\gamma) = \begin{bmatrix} \exp(\gamma) & 0 \\ 0 & \exp(-\gamma) \end{bmatrix}$$

*Equation 100*

then:

$$\mathbf{\Sigma} \equiv \kappa\, \mathbf{\Gamma}$$

*Equation 101*

Notably, like time reversal, the persymmetric transposition reverses the direction of propagation and hence the order of the matrices in a decomposition. Moreover, its action on the gain/loss matrix:

$$\mathbf{\Gamma} \xrightarrow{\mathcal{P}} \mathbf{J}\mathbf{\Gamma}^T\mathbf{J} = \begin{bmatrix} \exp(-\gamma) & 0 \\ 0 & \exp(\gamma) \end{bmatrix}$$

*Equation 102*

exchanges the gain and loss loops in agreement with time reversal.

When acting on a special unitary matrix $\mathbf{U} \in SU(2)$ the parity conjugation operation

$$\mathbf{U} \xrightarrow{\mathcal{P}} \mathbf{J}\mathbf{U}^T\mathbf{J}$$

*Equation 103*

and time reversal operation:

$$\mathbf{U} \xrightarrow{\mathcal{T}} \mathbf{U}^T$$

*Equation 104*

agree only if $\mathbf{U}$ is bisymmetric. Any $\mathbf{U} \in SU(2)$ can be expressed as the exponentiation of an anti-Hermitian matrix. The space of traceless Hermitian $2 \times 2$ matrices is spanned by the Pauli matrices:

$$\mathbf{J}_1 = \begin{bmatrix} 0 & 1 \\ 1 & 0 \end{bmatrix} \; ; \; \mathbf{J}_2 = \begin{bmatrix} 0 & -i \\ i & 0 \end{bmatrix} \; ; \; \mathbf{J}_3 = \begin{bmatrix} 1 & 0 \\ 0 & -1 \end{bmatrix}$$

*Equation 105*

of which only $\mathbf{J}_1 \equiv \mathbf{J}$ is bisymmetric. Consequently, there exists a unique one-parameter group of bisymmetric $\mathbf{U}(\theta) \in SU(2)$ given by:

$$\mathbf{U}(\theta) = \exp(i\theta\mathbf{J}_1) \equiv \begin{bmatrix} \cos(\theta) & i\sin(\theta) \\ i\sin(\theta) & \cos(\theta) \end{bmatrix}$$

*Equation 106*

with the useful property:

$$\mathbf{U}(\theta_1 + \theta_2) = \mathbf{U}(\theta_1)\mathbf{U}(\theta_2)$$

*Equation 107*

Moreover, $\mathbf{J}_1$ admits the eigenvalue-eigenvector decomposition

$$\mathbf{J}_1 = \mathbf{H}\mathbf{J}_3\mathbf{H} \; ; \; \mathbf{H} = \frac{1}{\sqrt{2}}\begin{bmatrix} 1 & 1 \\ 1 & -1 \end{bmatrix}$$

*Equation 108*

Consequently, $\mathbf{U}(\theta)$ admits the decomposition:

$$\mathbf{U}(\theta) = \mathbf{H}\exp(i\theta\mathbf{J}_3)\,\mathbf{H}$$

*Equation 109*

# $\mathcal{PT}-$symmetric cross-injection dual optoelectronic oscillator

which is recognised as describing the transmission of a MZI with a differential phase shift of $\theta$ its arms formed between two 180° hybrid couplers represented by the Hadamard matrices $\mathbf{H}$.

## $\mathcal{PT}-$ symmetry & the $SU(1,1)$ group

The analysis in the preceding demonstrates that a $\mathcal{PT}-$symmetric $2 \times 2$ coupler has a transmission matrix $\mathbf{K}$ that satisfies:

$$\mathbf{K} = \mathbf{J}_1 (\mathbf{K}^\dagger)^{-1} \mathbf{J}_1 \quad ; \quad \det(\mathbf{K}) = 1$$

*Equation 110*

and admits the decomposition:

$$\mathbf{K} = \mathbf{U}(\beta)\, \mathbf{\Gamma}(\gamma) \mathbf{U}(\alpha)$$

*Equation 111*

$$\mathbf{U}(\alpha) = \mathbf{H} \exp(i\alpha \mathbf{J}_3)\, \mathbf{H}$$

*Equation 112*

$$\mathbf{U}(\beta) = \mathbf{H} \exp(i\beta \mathbf{J}_3)\, \mathbf{H}$$

*Equation 113*

The merit of the decomposition is that it may be interpreted as a gain/loss block with diagonal transmission matrix $\mathbf{\Gamma}(\gamma)$ with gain/loss parameter $\gamma \in \mathbb{R}$ placed between two variable couplers each implemented by a MZI formed by a pair of 180° hybrid couplers with transmission matrix $\mathbf{H}$ connected by arms containing phase shifters providing a differential phase shift of $\alpha, \beta \in \mathbb{R}$.

Equation 110 may be re-expressed as:

$$\mathbf{K} \mathbf{J}_1 \mathbf{K}^\dagger = \mathbf{J}_1 \quad ; \quad \det(\mathbf{K}) = 1$$

*Equation 114*

Substituting:

$$\mathbf{K} = \mathbf{H} \mathbf{V} \mathbf{H}$$

*Equation 115*

and making use of Equation 108 yields:

$$\mathbf{V} \mathbf{J}_3 \mathbf{V}^\dagger = \mathbf{J}_3 \quad ; \quad \det(\mathbf{V}) = 1$$

*Equation 116*

The set of all $\mathbf{V} \in \mathbb{C}^{2\times 2}$ satisfying Equation 116 defines the group $SU(1,1)$. Consequently, the set of all $\mathcal{PT}-$symmetric coupler transmission matrices $\mathbf{K} \in \mathbb{C}^{2\times 2}$ is *isomorphic* to $SU(1,1)$.

Substituting the decomposition Equation 111, Equation 112, Equation 113 into Equation 115 and making use of Equation 108 yields:

$$\mathbf{V} = \exp(i\beta \mathbf{J}_3) \exp(\gamma \mathbf{J}_1) \exp(i\alpha \mathbf{J}_3)$$

*Equation 117*

Explicitly:

$$\exp(\gamma \mathbf{J}_1) = \begin{bmatrix} \cosh(\gamma) & \sinh(\gamma) \\ \sinh(\gamma) & \cosh(\gamma) \end{bmatrix} = \mathbf{H} \mathbf{\Gamma}(\gamma) \mathbf{H}$$

*Equation 118*

$$\exp(i\alpha \mathbf{J}_3) = \begin{bmatrix} \exp(i\alpha) & 0 \\ 0 & \exp(-i\alpha) \end{bmatrix}$$

*Equation 119*

$$\exp(i\beta \mathbf{J}_3) = \begin{bmatrix} \exp(i\beta) & 0 \\ 0 & \exp(-i\beta) \end{bmatrix}$$

# $\mathcal{PT}-$symmetric cross-injection dual optoelectronic oscillator

Consequently:

*Equation 120*

$$\mathbf{V} = \begin{bmatrix} \cosh(\gamma)\exp(i(\alpha+\beta)) & \sinh(\gamma)\exp(-i(\alpha-\beta)) \\ \sinh(\gamma)\exp(i(\alpha-\beta)) & \cosh(\gamma)\exp(-i(\alpha+\beta)) \end{bmatrix}$$

*Equation 121*

which is a general representation of an element of $SU(1,1)$ parameterised by $\alpha, \beta, \gamma \in \mathbb{R}$.

## 7. Appendix II Cross-injection dual oscillator

In this appendix a model of a TDO under injection introduced in reference [27] is first summarised introducing notation and then extended to a dual time delay oscillator which is shown to be $\mathcal{PT}-$symmetric.

The complex envelope $u$ representing the oscillation following the point of injection is the vector sum of the complex envelope $v$ representing oscillation prior at the point of injection and the complex envelope $w$ representing the injected carrier:

$$u = v + w \quad ; \quad \begin{cases} u = a_u \exp(i\theta_u) \\ v = a_v \exp(i\theta_v) \\ w = a_w \exp(i\theta_w) \end{cases}$$

*Equation 122*

where $a_u$, $a_v$, $a_w$ are the magnitude and $\theta_u$, $\theta_v$, $\theta_w$ the phase of the complex envelope $u$, $v$, $w$ respectively.

The evolution of the phase of the oscillation is described by a system of three equations corresponding to: the Leeson model of an RF filter with impulse response $h$ following the delay line Equation 123; the phase shift $\varphi$ induced by the injected carrier Equation 124; and the Barkhausen phase condition for oscillation Equation 125:

$$\theta_v = (h \otimes D_{\tau_D})\theta_u$$

*Equation 123*

$$\varphi = \tan^{-1}\left(\frac{\rho \sin(\theta_w - \theta_v)}{1 + \rho \cos(\theta_w - \theta_v)}\right)$$

*Equation 124*

$$\theta_u = \varphi + \phi + \theta_v \quad \mod 2\pi$$

*Equation 125*

where $\otimes$ denotes a convolution and $D_{\tau_D}$ represents the delay operator defined by:

$$(D_{\tau_D}u)(t) = u(t - \tau_D)$$

*Equation 126*

$\tau_D$ is the delay time, $\rho$ is the injection ratio:

$$\rho = a_w/a_v$$

*Equation 127*

and $\phi$ is the tuning phase bias.

In the special case of a single pole approximation of the low pass baseband equivalent model of a BPF filter, Equation 130 may be written:

# $\mathcal{PT}-$symmetric cross-injection dual optoelectronic oscillator

$$\tau_R \frac{d\theta_v}{dt} + \theta_v = D_{\tau_D}\theta_u$$

*Equation 128*

where $\tau_R$ is a time constant characterizing the bandwidth of the filter[3]. The system may be combined into a single recursive equation:

$$(I - h \otimes D_{\tau_D})\theta_v = (h \otimes D_{\tau_D})\left[\phi + \tan^{-1}\left(\frac{\rho \sin(\theta_w - \theta_v)}{1 + \rho \cos(\theta_w - \theta_v)}\right)\right]$$

*Equation 129*

Generalising the theory of a single injection locked oscillator to a dual TDO with cross-injection yields two coupled evolution equations for the phase $\theta_v = \theta_1, \theta_2$ of each individual oscillator subject to an injected carrier phase $\theta_w = \theta_2 + \xi_1, \theta_1 + \xi_2$ respectively:

$$\theta_1 = (h \otimes D_{\tau_1})\left[\phi_1 + \tan^{-1}\left(\frac{\rho_1 \sin(\xi_1 + \theta_2 - \theta_1)}{1 + \rho_1 \cos(\xi_1 + \theta_2 - \theta_1)}\right) + \theta_1\right]$$

*Equation 130*

$$\theta_2 = (h \otimes D_{\tau_2})\left[\phi_2 + \tan^{-1}\left(\frac{\rho_2 \sin(\xi_2 + \theta_1 - \theta_2)}{1 + \rho_2 \cos(\xi_2 + \theta_1 - \theta_2)}\right) + \theta_2\right]$$

*Equation 131*

where $\rho_1, \rho_2$ are the injection ratios; $\xi_1, \xi_2$ are the injection phase biases; $\tau_1, \tau_2$ are the delays and $\phi_1, \phi_2$ are the loop tuning phases of oscillator 1 and oscillator 2 respectively.

The effect of $\xi_1, \xi_2$ is to detune the figure of eight loop independently of the tuning $\phi_1, \phi_2$ of its component loops. Assuming the phase biases are time independent only the common part:

$$\xi_c = (\xi_1 + \xi_2)/2$$

*Equation 132*

plays a substantive role in the dynamics and the differential part:

$$\xi_d = (\xi_1 - \xi_2)/2$$

*Equation 133*

plays a minor role introducing an additive constant to the phase difference between the two oscillators. This constant may be distributed between $\theta_1, \theta_2$. For example, a quadrature phase oscillator, which corresponds to $\xi_1 = \pi/2, \xi_2 = -\pi/2$, has a solution that is the same as a dual oscillator with $\xi_1 = 0$, $\xi_2 = 0$ except for the addition of a constant $\pi/2$ phase difference that brings the oscillators into quadrature.

Consequently, one can restrict study to the system:

$$\theta_1 = (h_1 \otimes D_{\tau_1})\left[\phi_1 + \tan^{-1}\left(\frac{\rho_1 \sin(\xi + \theta_2 - \theta_1)}{1 + \rho_1 \cos(\xi + \theta_2 - \theta_1)}\right) + \theta_1\right]$$

*Equation 134*

$$\theta_2 = (h_2 \otimes D_{\tau_2})\left[\phi_2 + \tan^{-1}\left(\frac{\rho_2 \sin(\xi + \theta_1 - \theta_2)}{1 + \rho_2 \cos(\xi + \theta_1 - \theta_2)}\right) + \theta_2\right]$$

*Equation 135*

---

[3] The time constant $\tau_R$ is the group delay of the RF filter at the passband centre frequency.

# $\mathcal{PT}-$symmetric cross-injection dual optoelectronic oscillator

where for simplicity of notation the subscript on the common part of the injection phase biases has been dropped.

The convolutions by the impulse response of the BPFs may be approximated by the identity operator for oscillation frequencies well within the passband of the BPFs. Assuming evolution of the free oscillators to a single mode initial state, in the absence of injection:

$$\theta_1(t) - \theta_1(0) = \omega_1 t \quad ; \quad \theta_2(t) - \theta_2(0) = \omega_2 t$$
$$\Rightarrow$$
$$\phi_1 = \omega_1 \tau_1 \quad ; \quad \phi_2 = \omega_2 \tau_2$$

*Equation 136*

A steady locked state under cross injection is described by a solution of the form:

$$\theta_1(t) - \theta_1(0) = \theta_2(t) - \theta_2(0) = \omega_\infty t \quad ; \quad \theta_2(t) - \theta_1(t) = \theta_\infty$$

*Equation 137*

where $\omega_\infty$ and $\theta_\infty$ are respectively the asymptotic locked frequency and phase difference between the two oscillators. It follows that:

$$(\omega_\infty - \omega_1)\tau_1 = \tan^{-1}\left(\frac{\rho_1 \sin(\xi + \theta_\infty)}{1 + \rho_1 \cos(\xi + \theta_\infty)}\right)$$

*Equation 138*

$$(\omega_\infty - \omega_2)\tau_2 = \tan^{-1}\left(\frac{\rho_2 \sin(\xi - \theta_\infty)}{1 + \rho_2 \cos(\xi - \theta_\infty)}\right)$$

*Equation 139*

Eliminating $\omega_\infty$ provides a nontrivial trigonometrical necessary condition for a solution to exist. An assumption of small injection $\rho_1, \rho_2 \ll 1$ which is most often valid in practical applications results in a simplification of the necessary condition to:

$$(\omega_2 - \omega_1)\sqrt{\tau_1 \tau_2} = \left(\sqrt{\frac{\tau_2}{\tau_1}}\rho_1 + \sqrt{\frac{\tau_1}{\tau_2}}\rho_2\right)\cos(\xi)\sin(\theta_\infty) + \left(\sqrt{\frac{\tau_2}{\tau_1}}\rho_1 - \sqrt{\frac{\tau_1}{\tau_2}}\rho_2\right)\sin(\xi)\cos(\theta_\infty)$$

*Equation 140*

Equation 140 shows that the locking range is maximised by an injection phase bias of $\xi = 0, \pi$ and minimised by $\xi = \pm \pi/2$. The locking process depends upon episodes of constructive interference between the oscillating carrier and the injected carrier that reduce the saturated gain below the threshold for free oscillation allowing the injected carrier to capture the oscillator. In the case of a dual oscillator with a lossless coupler, which has a common cross-path phase shift of $\pm \pi/2$, an episode of constructive interference at one port corresponds to an episode of destructive interference at the other port. Locking is most effective when constructive interference occurs at both ports concurrently, which is the case for $\xi = 0, \pi$ and necessarily implies a coupler that does not conserve energy as is the case with the $SU(1,1)$ coupler considered in the main text. It is clear that there is latitude in the precision parameter values can be set, and while not critical best performance is achieved if attention is paid to matching the through- and cross- paths so that $\xi \sim 0$ and ensuring that the detuning $\omega_2 - \omega_1$ of the desired oscillating resonance of each oscillator is minimal so that $\theta_\infty \sim 0$ as is assumed in the main text.

In the main text a mode-selective $\mathcal{PT} -$ symmetric dual TDO circuit architecture has been derived from first principles and shown to be equivalent to a symmetric cross-injection dual TDO with Vernier effect mode-selection provided by similar but distinct time delays. The logic of the argument may be reversed by starting with the cross-injection dual TDO and showing that it is equivalent to a $\mathcal{PT} -$ symmetric dual TDO.

# $\mathcal{PT}-$symmetric cross-injection dual optoelectronic oscillator

Consider the cross-injection coupler transmission matrix:

$$\mathbf{P} = \begin{bmatrix} 1 & \rho \\ \rho & 1 \end{bmatrix}$$

*Equation 141*

The eigenvalues and associated eigenvectors of $\mathbf{P}$ are given by:

$$\lambda_\pm = 1 \pm \rho \quad ; \quad \boldsymbol{u}_\pm = \frac{1}{\sqrt{2}} \begin{bmatrix} 1 \\ \pm 1 \end{bmatrix}$$

*Equation 142*

Hence $\mathbf{P}$ may be written:

$$\mathbf{P} = \sqrt{1-\rho^2}\,\mathbf{H}\boldsymbol{\Gamma}\mathbf{H}$$

*Equation 143*

where:

$$\mathbf{H} = \frac{1}{\sqrt{2}} \begin{bmatrix} 1 & 1 \\ 1 & -1 \end{bmatrix} \quad ; \quad \boldsymbol{\Gamma} = \begin{bmatrix} \exp(\gamma) & 0 \\ 0 & \exp(-\gamma) \end{bmatrix} \quad ; \quad \exp(\gamma) = \sqrt{\frac{1+\rho}{1-\rho}} \Rightarrow \rho = \tanh(\gamma)$$

*Equation 144*

and interpreted as describing a gain/loss block with transmission matrix $\boldsymbol{\Gamma}$ placed between two Hadamard couplers (i.e., 180° hybrid couplers) with transmission matrix $\mathbf{H}$. Provided $\rho^2 \neq 1$, the scalar $\sqrt{1-\rho^2}$ may be absorbed into the overall gain control parameter $\kappa$.

Within the complete dual ring structure of the oscillator, the two Hadamard couplers form a MZI containing the two delay lines with delay time $\tau_1$, $\tau_2$ together with the limiting RF amplifiers within its arms. One may decompose this MZI into the concatenation of a dual matched delay line with delay time $\tau_2$ and an MZI with the limiting amplifiers within both its arms but containing a delay with delay time $\tau_1 - \tau_2$ in one arm only (see main text Figure 4). Replacing the two limiting RF amplifiers by a $U(2)-$invariant dual RF amplifier permits its removal from the MZI and placement adjacent to the gain/loss block, thereby leading to the original mode-selective $\mathcal{PT}-$ symmetric dual TDO circuit architecture.

The intermediate result where the cross-injection coupler is replaced by a gain/loss block placed between two Hadamard couplers provides new insights into the cross-injection dual oscillator architecture. Effective injection locking requires successive episodes of destructive interference between the injected carrier and the oscillation that reduce the saturated gain below the threshold for free oscillation allowing the injected carrier to capture the oscillator. When one output port of a $2 \times 2$ *lossless* coupler experiences destructive interference the other port necessarily experiences constructive interference which results in marginal locking of a cross-injection dual oscillator. This is the physical reason why the cross-injection coupler must have a real transmission matrix up to a scalar multiplier. Given that the coupler thereby is not energy conserving, it is inevitable that the diagonalization of its transmission matrix normalised to unit determinant will result in a gain/loss block. This is true even for asymmetric cross-injection for which the gain/loss is determined by the geometric mean $\sqrt{\rho_1 \rho_2}$. However, the eigenvectors are then no longer orthogonal having inner product:

$$(\boldsymbol{u}_+, \boldsymbol{u}_-) = \frac{\rho_1 - \rho_2}{\rho_1 + \rho_2}$$

*Equation 145*

and the matrix $[\boldsymbol{u}_+ \ \boldsymbol{u}_-]$ and its inverse depart from a Hadamard matrix.

It can be understood from the $\mathcal{PT}-$symmetry perspective that the two RF amplifiers must have matched characteristics to avoid undesired imbalance of the MZI. Strictly, this requires a $U(2)-$invariant dual RF amplifier. In the case of two independent RF amplifiers in the linear regime, to avoid imbalance, it is sufficient that the amplifiers have matched linear gains. Moreover, once the loss-path oscillation has decayed to negligible proportions (i.e., dark-port extinction is complete) the magnitude of the oscillation in each arm of the MZI is comparable (equal for Hadamard couplers) and it is sufficient that the amplifiers have matched saturated gain. Transient, imbalance remains possible in the brief interval between the linear and the dark-port extinction regimes. However, observations suggest this is not an issue.